\documentstyle[emulateapj]{article}

\def\kms{km\,s$^{-1}$\,}
\def\vs{$V_{S}$\,\,}
\def\vb{$V_{B}$\,\,}
\def\etal{ et~al.\rm\,}

\def\fl{ergs cm$^{-2}$ s$^{-1}$}
\def\tetp{$(T_e/T_p)_0$}
\def\totp{$(T_O/T_p)_0$}
\def\fuse{{\it FUSE}}
\def\chan{{\it Chandra}}
\def\eins{{\it Einstein}}

\def\dem71{DEM\,L\,71}
\def\onine{SNR\,0509$-$67.5}
\def\lybovi{F(Ly $\beta$)/F$_{O\,VI}$(1032)}
\def\onenine{SNR\,0519$-$69.0}
\def\foureight{SNR\,0548$-$70.4}

\begin{document}

\title{The Detection of Far Ultraviolet Line Emission from Balmer-Dominated Supernova Remnants
in the Large Magellanic Cloud\altaffilmark{1} }

\author{Parviz Ghavamian\altaffilmark{2}, William P. Blair\altaffilmark{2}, Ravi
Sankrit\altaffilmark{3}, John C. Raymond\altaffilmark{4} and John P. Hughes\altaffilmark{5}}

\altaffiltext{1}{This work is based on data obtained for the Guaranteed Time Team by the NASA-CNES-CSA
FUSE mission operated by the Johns Hopkins University. Financial support to U.S.
participants has been provided by NASA contract NAS5-32985.}
\altaffiltext{2}{Department of Physics and Astronomy, Johns Hopkins University, 3400 North
Charles Street, Baltimore, MD, 21218-2686}
\altaffiltext{3}{Space Science Laboratory, University of California, Berkeley, Berkeley, CA 94270-7450}
\altaffiltext{4}{Harvard-Smithsonian Center for Astrophysics, 60 Garden Street, Cambridge, MA, 02138}
\altaffiltext{5}{Department of Physics and Astronomy, Rutgers University, Piscataway, NJ, 08854-8019}

\begin{abstract}

We present the first far ultraviolet (FUV) spectra of the four known Balmer-dominated
supernova remnants (SNRs) in the Large Magellanic Cloud, acquired with the Far Ultraviolet Spectroscopic
Explorer (\fuse).  The remnants \dem71 (SNR 0505$-$67.9), \onine, \onenine\, and \foureight\,
are all in the non-radiative stages
of evolution and exhibit expansion speeds ranging from $\sim$ 500 \kms to $\sim$ 5000 \kms.
We have detected broad emission
lines of Ly $\beta$, Ly $\gamma$, C~III and O~VI in \dem71\, ($V_{FWHM}$ $\sim$\,1000 \kms) and
have detected broad Ly $\beta$ and O~VI emission in \onenine\, ($V_{FWHM}$ $\sim$\,3000 \kms). 
In addition, broad Ly $\beta$ emission ($V_{FWHM}$ $\sim$\,3700 \kms) has been observed in \onine,
the first detection of broad line emission from this SNR.
No emission was detected in our \fuse\, spectrum of \foureight, allowing us to place only upper limits on 
the FUV line fluxes.  The spectra of these SNRs are unaffected by postshock cooling, and provide
valuable probes of collisionless heating efficiency in high Mach number shocks.   We have used the 
\lybovi\, flux ratio and relative widths of the
broad Ly $\beta$ and O~VI lines to estimate the degree of electron-proton and
proton-oxygen ion equilibration in \dem71, \onine\, and \onenine.  Although
our equilibration estimates are subject to considerable uncertainty due to
the faintness of the FUV lines and contributions from bulk Doppler broadening, our 
results are consistent with a declining efficiency
of electron-proton and proton-oxygen ion equilibration with increasing shock speed.  From
our shock velocity estimates we obtain ages of 295-585 years for \onine\, and 520-900 years
for \onenine, respectively, in good agreement with the ages obtained from
SN light echo studies.

\end{abstract}

\section{INTRODUCTION}

The proximity of the Large Magellanic Cloud 
(LMC) (50 kpc; Feast 1991, 1999) and the relatively small extinction along
the line of sight to this galaxy (E(B-V) $\sim$ 0.1) make it an excellent laboratory
for study in the X-ray and ultraviolet.  In particular, the well calibrated distance of
the LMC allows us to study both the kinematics of individual supernova
remnants (SNRs) (Blair \etal\, 2006) and properties of the global SNR distribution within this galaxy.

The first systematic X-ray survey of the LMC was performed with the Imaging Proportional
Counter of the \eins\, observatory (Long, Helfand \& Grabelsky 1981) in the 0.15-4.5 keV
range.  The observations revealed that nearly half of the 97 detected sources exhibited
an extended morphology characteristic of SNRs, making these objects the largest class
of X-ray sources revealed in the survey.   Four of the SNRs identified with \eins\, were
shown by Tuohy \etal\, (1982) to exhibit optical spectra consisting
almost entirely of faint hydrogen Balmer line emission (i.e., they are `Balmer-dominated').  The four
remnants $-$ 0505$-$67.9 (DEM L 71 from the Davies, Elliott \& Meaburn (1971) catalogue), \onine,
\onenine\, and \foureight\, vary widely in age, ranging from a mere few hundred years
(\onine; Rest \etal\, 2005) to $\sim$10,000 years (\foureight, Smith \etal\, 1991).

Balmer-dominated line emission is produced when the expanding SNR
encounters low density ($\sim\,$0.1-1 cm$^{-3}$), partially neutral interstellar
ISM (Chevalier \& Raymond 1978; Bychkov \& Lebedev 1979).  Due to the high temperature and
low density behind the blast wave, the cooling time scale of the gas greatly exceeds
the dynamical time scale of these SNRs (they are non-radiative).  The hydrogen overrun by the shock is collisionally
excited in a thin ($\lesssim$10$^{15}$ cm) ionization zone, producing optical spectra dominated
by the Balmer lines of hydrogen (Chevalier \& Raymond 1978; Chevalier, Kirshner \& Raymond
1980).  In the ultraviolet, the spectra are
'Lyman-dominated'.  Balmer-dominated shocks are mostly found in young remnants (e.g., SNRs in the
Sedov-Taylor stage or younger) of Type Ia explosions.  In these objects the survival
of significantly neutral, low density ISM gas is likely aided by the lack of 
strong ionizing radiation and stellar wind from the progenitor star (a white dwarf).
X-ray spectroscopy of the reverse-shocked ejecta
in the four LMC Balmer-dominated remnants has revealed strong Fe L shell emission 
(Hughes \etal\, 2003; Hendrick, Borkowski \& Reynolds 2003; Warren \& Hughes 2004),
confirming the Type Ia origin of these SNRs.
These overall characteristics place the four Balmer-dominated LMC remnants into
the same class as the well-known Galactic Type Ia SNRs Tycho and SN 1006 (Chevalier, Kirshner \& Raymond
1980; Kirshner, Winkler \& Chevalier 1987; Smith \etal\, 1991; Ghavamian \etal\, 2001),
making them the first examples of such remnants identified outside our galaxy.  

The hydrogen line emission in Balmer-dominated shocks consists of two components:
a narrow component produced by collisional excitation of overrun ambient H~I, and
a broad component generated when the hot postshock H~I produced by charge exchange is collisionally
excited.  Both broad and narrow component Balmer lines can often be detected
in the optical, but interstellar absorption in the ultraviolet completely removes the
narrow component Lyman line emission.   However, if the broad component Lyman lines are both
bright and wide, the profile wings can be detected.  

The diagnostic utility of the broad H~I lines is twofold.  First, the width of
a broad H~I line gives the postshock proton temperature, thus yielding an estimate of
the shock speed when combined with the Rankine-Hugoniot jump conditions (Chevalier, Kirshner
\& Raymond 1980; Smith \etal\, 1991; Ghavamian \etal\, 2001).  Second, comparing
the width and flux of the Lyman lines with those of ultraviolet resonance lines such as 
O~VI $\lambda\lambda$1032, 1038 and C~III $\lambda$977 yields an estimate of the ion-ion temperature
equilibration at the shock front, well before the equipartition of energy between the
different ionic species by Coulomb collisions (Hester, Raymond \& Blair 1994; Raymond, Blair \& Long 1995). 
This is key information for probing the nature of collisionless (plasma) heating processes at the
shock front, one of the central problems in understanding high Mach number shocks in
many astrophysical settings.

The availability of broad hydrogen, carbon and oxygen emission lines,
along with high quality X-ray spectra from \chan\,
provides us with an invaluable opportunity to simultaneously study both the electron-proton and ion-ion
equilibration in fast collisionless shocks and the evolution of remnants from Type Ia SNe.  In this
paper we present FUV spectra of the four Balmer-dominated LMC SNRs.  We utilize numerical shock
models to interpret the spectra, constrain the kinematics of the four SNRs and investigate collisionless 
heating processes in fast astrophysical shock waves.

\section{OBSERVATIONS AND DATA REDUCTION}

\subsection{THE RAW DATA}

We observed the four Balmer-dominated LMC SNRs with the Far Ultraviolet Spectroscopic Explorer
(\fuse) in 2001 September 23 and 24 (see Moos \etal\, 2000 and Sahnow \etal\, 2000 for a detailed
description of the spectrograph).  The data were acquired as part of a Cycle 3 Guaranteed Time Observation
program (P214; W. P. Blair, PI), with each remnant observed through the low resolution (LWRS)
square aperture of \fuse\, (30\arcsec\,$\times$\,30\arcsec).  In each case the LWRS aperture
was centered on the geometric center of each SNR as measured in the optical by Tuohy \etal\, (1982).
The roll angles of these observations were set by the nominal spacecraft roll at the time of data
acquisition.  The total integration time and the time obtained during orbital night
is listed for each object in Table~1.
The Balmer-dominated LMC SNRs vary in size from approximately 30\arcsec\,
across (\onine) to 1\farcm1 across (\foureight), resulting in significant variations
in the fraction of the shell covered in each observation (see Figure~1).  While the LWRS aperture
covered nearly the entire shell of \onine\, and \onenine, it sampled primarily 
the approaching and receding faces of the shell in \dem71 and \foureight.  As we show
in the following sections, the fraction of each SNR covered by the \fuse\, aperture can significantly
affect the shapes and centroids of the FUV lines in the final extracted spectra.

We acquired a separate \fuse\, observation of DEM L71 during Cycle 3 (Guest Observer program
C072; PI: P. Ghavamian).  In this observation we placed the MDRS aperture (4\arcsec\,$\times$20\arcsec)
parallel to the southern edge of the blast wave at a position angle of 113$^{\circ}$ (see Figure~1).  The
aperture orientations were nearly constant during these observations.   A total
integration time of 121.6 ks was accumulated during five separate pointings 
on 2003 January 4-7 (C0720202-C0720205) and 2003 December 29 (C0720206).  However, the target
acquisition in observation C0720205 failed, leaving 100 ks of useful data.  

Our data reduction was performed using version 3.1.3 of the CalFUSE pipeline.  The expected
FUV line emission in the four LMC remnants is expected to be faint ($\lesssim$10$^{-13}$
ergs cm$^{-2}$ s$^{-1}$), with line widths ($\gtrsim$1000 \kms) greatly exceeding the resolution
of the filled LWRS aperture (approximately 106 \kms).  These properties make the detection of the FUV lines
quite sensitive to the quality of the background subtraction.  Since the CalFUSE pipeline derives an
empirical background for each spectrum using non-illuminated portions of the detector, we can
improve the signal-to-noise of the measured background by co-adding the individual raw exposures for
each of the four FUSE detectors (1A,1B,2A,2B), then extracting 1-D spectra from the combined raw frames.
We used the IDL routine {\sf ttag\_combine} to combine the raw exposures
(Figure~2), then used the CalFUSE pipeline to extract the 1-D spectra.  We retained only
the orbital night portions from each dataset (Table~1) to minimize the contribution of
airglow emission lines and scattered airglow continuum to the object spectra.

We combined the raw exposures from the MDRS data in the same manner described above for the LWRS
data. Since the C0720206 data were acquired nearly a year after the other
observations, the FUV spectrum of this observation is significantly shifted in
detector pixels relative to the earlier observations.  Therefore, we first co-added 
the C0720202-C0720204 datasets into one raw image, then co-added
C0720206 into a separate raw image.  We extracted spectra from each image, 
cross-correlated each one and finally co-added them to obtain the
1-D MDRS spectrum of the DEM L71 blast wave.  

\subsection{SPECTRAL EXTRACTION}

The FUSE spectra are recorded on four channels: LiF 1, LiF2, SiC 1 and SiC 2.  Each channel
is divided into A and B segments, with the two segments covering the approximate wavelength range
905 \AA\,-\,1100 \AA\, for each of the LiF 1 and LiF 2 channels, and the wavelength range
987 \AA\,-\, 1180 \AA\, for the SiC 1 and SiC 2 channels (see the FUSE Observer's Guide\footnote{
http://fuse.pha.jhu.edu/support/guide/guide.html}).
In each of our LWRS observations, the LiF 1A and SiC 2A (917 \AA\,-\,1006 \AA)
segments offer the highest effective area at the positions of the expected FUV emission lines.
Therefore, we searched the spectra from these segments for SNR emission first.  Although the
extracted \fuse\, spectra are meant to be summed into one high signal-to-noise
data product, inter-channel thermal drifts during each exposure can result in 
misalignments as large as 10\arcsec\, between LiF 1 and the other channels.  
While the LiF 1 channel was held fixed on the SNR during an observation, the others
sampled emission from slightly different portions of the SNR.  The resulting
spectra can differ from channel to channel.  Therefore,
for each observed SNR we combined spectra from multiple channels only when they appeared consistent with 
one another to within the flux uncertainties.

In the case of \dem71, prominent, broad emission lines of Ly $\beta$, Ly $\gamma$, O~VI
and C~III are detected in the LWRS observations.   The spectral features are observed in all
segments of the \fuse\, detector save for the LiF 1B and LiF 2A segments, whose spectral range
excludes these emission lines. Although inter-channel drift is expected to
have occurred during the 28 ks observation, the spectra are all consistent with one another
to within the errors.  We generated the spectra in Figure~3 by cross-correlating and
combining data from SiC 2A with SiC 1B (905 \AA\,$-$\,1005 \AA) and the data from
LiF 1A with those of LiF 2A and SiC 1A (988 \AA\,$-$\,1090 \AA).  Although
broad Ly $\beta$ and O~VI emission were seen in the SiC 2B segment, we did not include
this data in our co-addition because the background was unevenly 
over-subtracted for this segment by the CalFUSE pipeline.

Unlike \dem71, \onine\, and \onenine\, fit almost completely within
the LWRS aperture (Figure~1), increasing the likelihood that inter-channel drifts
would adversely affect the detection of FUV emission from these objects.  Since \onine\, exhibits
the faintest FUV emission of all the Balmer-dominated SNRs (Figure~4), we
restricted our analysis of this object to data from the LiF 1A segment.

In the case of SNR 0519$-$69.0, on the other hand, FUV line emission was detected
in all segments shortward of 1100 \AA.  One factor complicating our interpretation
is the presence of a very faint continuum under the Ly $\beta$
and O~VI features (clearly seen in Figure~2).  The continuum is absent on the
short wavelength side of the broad Ly $\beta$ line, but rises to a nearly constant level 
beneath the broad lines and continues to the long wavelength end of the spectrum.
The continuum is absent in the other channel spectra near 1030 \AA, and appears
to be stellar contamination.  After comparing the emission in all channels,
we cross-correlated and combined the LiF 1A, LiF 2B, SiC 1A and SiC 2B 
data to obtain spectra in the 940 \AA\,$-$1050 \AA\, range.  Although broad
Ly $\gamma$ was also detected, it was present solely in the SiC 2A
spectrum of \onenine\, (the channel with the highest effective area
in this wavelength region).

In our MDRS observation of \dem71, FUV emission was detected only
in the spectrum of the LiF 1A segment.  The lack of detection in the other segments is likely
due to a combination of their lower effective areas and the 
the strong impact of inter-channel thermal drifts.
The latter effect is especially important here, where the
smaller size of the MDRS aperture makes it easier for non-primary channels to drift
off the targeted filament.

One of the more surprising results of our observations is the lack of any
discernible FUV emission from the largest of the Balmer-dominated SNRs,
\foureight\, (see Figure~6).  No emission is detected in either the raw combined data,
the individual channel spectra or the co-added spectra from all channels.  
In the following section we will describe a possible explanation for this
non-detection, and discuss its implications.

\section{FAR ULTRAVIOLET SPECTRA}

Of the observed Balmer-dominated remnants, two objects $-$ \dem71\, and \onenine\, $-$
exhibit readily identifiable, broad FUV line emission consistent with Balmer-dominated
shocks: Ly $\beta$ 
($\lambda$1025.7), O~VI $\lambda\lambda$1032, 1038, Ly $\gamma$ ($\lambda$972.5) and
C~III $\lambda$977.  In a third remnant, \onine, only the broad Ly $\beta$ line
is detected with reasonable certainty.  However, this is the first detection of broad
hydrogen line emission from this remnant.  Surprisingly, no discernible FUV emission is
seen from the fourth remnant, \foureight, despite the known detection
of broad Balmer line emission in this remnant by Smith \etal\, (1991).   Here, we describe 
the \fuse\, spectra of each remnant in detail.  All spectral fits have been performed using
the {\sf SPECFIT} task (Kriss 1994) from the {\sf STSDAS} contributed
package in IRAF\footnote{IRAF is distributed by the National Optical Astronomy
Observatories, which is operated by the AURA, Inc. under cooperative agreement with the National Science Foundation.}
unless noted otherwise.

\subsection{\dem71}

\subsubsection{LWRS Spectra}

The combined LWRS spectrum of \dem71\, (Figure~3) provides an especially impressive example of
the FUV emission from a Balmer-dominated SNR.  The emission lines are very broad
($\sim$1000 \kms), bright and clearly impacted by Galactic and LMC halo absorption.  The
presence of strong, broad Ly $\beta$ indicates that a significant fraction of the gas 
entering the blast wave is neutral, while the relative faintness of the C~III line relative
to O~VI indicates that postshock cooling is negligible (Raymond 2001).  The
wavelengths of prominent absorption features from both the Milky Way and LMC are marked in Figure~3.
The broad Ly $\beta$, O~VI $\lambda$1032 and O~VI $\lambda$1038 lines are moderately
blended, as are the Ly $\gamma$ and C~III $\lambda$977 lines. 

The LWRS aperture samples emission near the projected center of \dem71\,
(Figure~2).  Therefore, each emission line should appear 
doubled by the bulk motion of the postshock gas ($\pm\,\frac{3}{4}\,v_{s}$) on near
and far sides of the SNR.  The shock velocities in \dem71\, have been
estimated to lie between 600 \kms\, and 1100 \kms\, (Smith \etal\, 1991; Ghavamian \etal\,
2003; Rakowski, Ghavamian \& Hughes 2003), so the emission lines in \dem71\, should
appear blue/red-shifted by 400$-$800 \kms\, from the rest velocity of the LMC (which
we take to be +275 \kms).  However, the presence of strong interstellar absorption near the center
of each emission line in Figure~3 makes it difficult to distinguish whether each
emission line is doubled.  As we show in Section~4, the line profile shapes
observed in \dem71\, are in fact consistent with the presence of two Doppler-shifted
components.  We have devised a spherical shell model of the blast wave to 
address the significant interplay between bulk and thermal Doppler
motions in this spectrum.  We present this model in the next section.

Although a spherical shell model is required for a kinematic interpretation
of the spectrum in Figure~3, we can still estimate
the observed flux in each of the broad emission features by fitting each with a
single Gaussian line.  Prior to fitting each spectrum we 
estimated the baseline level by fitting the data on the blue
and red ends of the spectrum (devoid of emission features) with a straight line
of zero slope.  After obtaining the best fit value to the baseline, we held
this quantity fixed during our subsequent fits to the emission lines.  We chose a 9-component fit
for the \dem71\, spectrum, fitting the emission components with
a flat baseline and three Gaussian emission lines for broad Ly $\beta$ and O~VI $\lambda\lambda$1032, 1038.
We included three absorption components for broad Ly $\beta$ (Galactic and LMC halo H~I absorption, Galactic
O~I absorption at 1027 \AA) while excluding the wavelength range covering the Ly $\beta$ airglow
line.  We also included one O~VI LMC absorption component for each of the
broad O~VI lines.  Although there is possible evidence for more absorption components along 
the line of sight, we found that adding more of these components did not substantially
improve the statistical quality of the overall fit.  

We improved the constraining power 
of our fits by tying together the widths and velocity centroids of the O~VI
$\lambda\lambda$1032, 1038 emission lines.  We applied the same constraint to
the O~VI absorption features, with the additional requirement $\tau_{1032}$\,:\,$\tau_{1038}$ = 2:1.  In fitting the emission
near 972 \AA\, we used a 5 component model consisting of a linear baseline, one Gaussian each for
Ly $\gamma$ and C~III $\lambda$977, and one Galactic and one LMC H~I absorption component.  
We held the velocity width and line centroid of the Ly $\gamma$ line equal to the values measured
from our Ly $\beta$ profile fit. 
We included the H~I absorption components solely as an aid for better fitting of the broad Ly $\gamma$ emission
line.  The fitted H~I absorption lines do not actually provide reliable information on the H~I column
density because the data quality near the cores of the absorption profiles are heavily compromised
by the bright airglow Ly $\beta$ line.  Furthermore, we used a Gaussian profile shape
to describe what are actually saturated H~I absorption lines, further invalidating a physical interpretation 
of the fitted parameters for these components.

The resulting emission line fit for \dem71\, is marked in Figure~3 and the predicted emission
line parameters are listed in Table~2.  The Ly $\beta$ line width (1135$\pm$30 \kms) is significantly
larger than that of the O~VI $\lambda$ 1032 line (740$\pm$45 \kms), already suggesting
that the protons and oxygen ions have undergone significant temperature equilibration at the
shock front.  However, we defer the task of estimating the specific shock velocities and 
comparing the results to existing optical measurements to Section~4.  We note the slight
velocity shift for the fitted emission line centroids: taking a systemic velocity of
+275 \kms\, for the LMC, the Ly $\beta$ and O~VI $\lambda$1032 lines are redshifted
by 100$-$150 \kms\, (Table~2).  This feature suggests that the emission from the far side of the
\dem71\, shell is slightly brighter than the near side.  In Section~4 we quantify this
result with more detailed. models of the blast wave shell.

The prominent absorption features present in the O~VI $\lambda\lambda$1032, 1038 lines
are centered at 250$\pm$50 \kms, indicating that they arise in the LMC halo.  The Galactic
absorption component is considerably weaker.  Our model fits indicate $\tau_{1032}$(Gal)\,=\,
1.7$\pm$0.6, with $\sigma\,=\,$50$\pm$22 \kms\, after deconvolving the measured FWHM with the filled slit
spectral response of 106 \kms.  The implied O~VI column density is approximately
(3.5$\pm$2.2)$\times$10$^{14}$ cm$^{-2}$.  Both the line width and column density lie at
the upper end of, but are fully 
consistent with, values measured by Howk \etal\, (2002) in their analysis of \fuse\,
spectra of 12 early-type stars in the LMC.

Aside from the O~VI doublet, C~III $\lambda$977 in \dem71\, is the only other heavy ion line detected
in our LWRS spectra of the Balmer-dominated LMC remnants.  Our single Gaussian fit (Table~2)
indicates that the C~III line is less than half as broad as the Ly $\beta$ line, indicating
that significant temperature equilibration has taken place between protons and carbon ions 
at the shock front.

\subsubsection{MDRS Spectrum}

The night-only MDRS spectrum of \dem71\, (Figure~7) reveals prominent, broad Ly $\beta$ emission
moderately blended with faint O~VI lines.  As with the face-on LWRS spectrum, we fit the
three lines together while including a fixed baseline level and both Galactic and LMC H~I
absorption.  The O~VI lines are faint enough to make the Galactic and LMC halo
absorption features undiscernible, so we did not include absorption in our O~VI profile
fits.  The inclusion of absorption components had minimal impact on the quality of the fit.
The final results, quoted in Table~2, indicate a broad Ly $\beta$ width
of 1365$\pm$75 \kms\, FWHM.  The O~VI line is significantly narrower, 935$\pm$125 \kms,
an indication that partial ion-ion equilibration has taken place at the shock transition.

\subsection{\onine}

In their optical spectroscopy of this SNR Tuohy \etal\, (1982) and Smith \etal\,
(1991) were unable to detect broad Balmer line emission and concluded that the line was
too faint and broad to be detected.   Our LWRS spectrum of \onine\, reveals a faint,
strongly blueshifted broad Ly $\beta$ line (Figure~4), the first broad emission line detected 
from this SNR to date.  The detected line emission is indeed faint,
amounting to only 20\% and 35\% of the broad Ly $\beta$ flux
in \dem71\, and \onenine, respectively.  We fit the broad Ly $\beta$ profile
with a single Gaussian and ignored the regions containing the Ly $\beta$
airglow and H~I absorption features.  The result of this fit (Table~2) is a broad
Ly $\beta$ width of 3710$\pm$400 \kms, where the uncertainty includes both the
random statistical error ($\pm$325 \kms) and the baseline uncertainty ($\pm$240
\kms; this quantity is the variation in line width when the background level
is fixed to the upper or lower limits allowed by the initial baseline fit).  This is the largest
H~I line width measured for a broad component in any of the Balmer-dominated SNRs
thus far.  

Aside from being very broad, the Ly $\beta$ line is blueshifted to $-$263$\pm$125 \kms, 
indicating a large velocity shift ($-$540 \kms) relative to the LMC frame.  The
velocity shift of the broad line center is proportional to the projected component
of the bulk velocity along the line of sight.  The blueshift of the line indicates
that the Ly $\beta$ emission arises on the near side of \onine.  As seen
in Figure~1, there is a prominent patch of enhanced H$\alpha$ emission stretching
inward from the southwest corner of the shell.  It is likely that this patch lies
on the near side of \onine\, and that \fuse\, has detected the FUV emission from this 
region.  As we will show in Section~4, the size of the Ly $\beta$ broad component shift
is fully consistent with the position of the patchy enhancement relative to the
geometric center of \onine.  Note that the region of enhanced emission
is isolated close to the edge of the shell, so the range of viewing angles to
this section of the blast wave is likely small (i.e., the emitting region
can be approximated as a plane-parallel shock tilted slightly toward 
the observer).  

\subsection{\onenine}

This Balmer-dominated SNR is similar in size to \onine\, (and likely of comparable
age as measured from light echoes from the SN explosion, Rest \etal\, 2005).  The
FUV spectrum (Figure~5) exhibits broad Ly $\beta$ emission line at 1026 \AA, blended with
broad O~VI $\lambda\lambda$1032, 1038 line emission. A broad Ly $\gamma$
line can clearly be discerned near 972 \AA\, although the C~III line falls below the
detection threshold.

We can draw a direct comparison between the broad Ly $\beta$ widths measured in 
\onenine\, and the H$\alpha$ widths measured by Tuohy \etal\, (1982) and
Smith \etal\, (1991).  After deconvolving with the filled slit resolution (106 \kms)
we find good agreement between the broad Ly $\beta$ width of 3130$\pm$155 \kms\,
and the H$\alpha$ value of 2800$\pm$300 \kms\, measured along the rim by
Tuohy \etal\, (1982).  However, our line width is over twice the 1300$\pm$200 \kms\, width 
measured for the broad H$\alpha$ line by Smith \etal\, (1991).  The reason for the difference
is that the spectroscopic slit used by Smith \etal\, (1991) was placed E-W across
\onenine, intersecting the bright knot on the eastern rim 
of the remnant (clearly seen in Figure~2).  The spectra of Smith \etal\, (1991)
were only deep enough to reveal broad H$\alpha$ emission from this bright knot,
where the blast wave has encountered a relatively dense region of the ISM and slowed.
On the other hand, Tuohy \etal\, (1982) observed the fainter, more uniform
portions of the \onenine\, rim, where considerably less deceleration has
occurred.  As shown in Figure~2, the position and orientation of the LWRS aperture
placed the brightened emission knots entirely outside the slit during the \fuse\, observation.
Therefore, our FUV spectrum samples the less decelerated portions of \onenine\, and
should give Ly $\beta$ widths closer to the results of Tuohy \etal\, (1982).

\subsection{\foureight}

There are no emission lines detected in the \fuse\, spectrum of this object (Figure~6).  
The broad H$\alpha$ width measured by Smith \etal\, (1991) along the eastern rim of \foureight\,
is 760$\pm$140 \kms, corresponding to shock speeds of 700$-$950 \kms.  Since we easily detected shocks of
similar speed in our LWRS \fuse\, spectrum of \dem71 in both Ly $\beta$ and O~VI, it appears that the 
blast wave region covered by our LWRS slit (Figure~1) is propagating into a very low density, mostly ionized
medium.  In that case, pressure conservation would require very high shock speeds at this position, meaning
that aside from being faint, the lines are probably also very broad.  

Interestingly, our \fuse\, aperture
includes several bright clumps of shocked interstellar material near the projected center
of \foureight\, (Figure~1). The lack of C~III and O~VI emission from the knots in our \fuse\, spectrum
suggests that they are excited by slow radiative shocks ($\lesssim$90 \kms; Hartigan, Raymond
\& Hartmann 1987).   In their narrowband 
imagery of \foureight\, Tuohy \etal\, (1982) detected all of these clumps in H$\alpha$, but
only detected [O~III] emission from the clumps lying outside the \fuse\, aperture.  In addition,
the \chan\, ACIS S image of \foureight\, (Hendrick, Borkowski \& Reynolds 2003) does not show any
obvious X-ray emission from clumps within the \fuse\, aperture.  These features are
all characteristic of slow radiative shocks and suggest that dense, localized material
has been overrun by the blast wave on the near and/or far side of \foureight.

\section{TEMPERATURE EQUILIBRATION AND KINEMATICS}

The \fuse\, spectra of the Balmer-dominated LMC SNRs are particularly valuable in that they feature
prominent emission lines of hydrogen, carbon and oxygen, allowing us to directly
measure the degree of ion-ion equilibration in fast non-radiative shocks.  The diagnostic
information available to us includes (1) the relative widths of the C~III $\lambda$977, Ly $\beta$
and O~VI $\lambda\lambda$1032, 1038 lines, which reflect both the relative temperatures
of these ions and the shock speed (assuming we know the contribution of bulk Doppler
broadening to the line profiles), and (2) the Ly $\beta$ / O~VI flux ratio, which
reflects the electron-proton equilibration, shock speed and preshock fraction of neutral
hydrogen.  

In the case of \onenine, the good agreement between our measured broad Ly $\beta$
width and the H$\alpha$ width measured by Tuohy \etal\, (1982) from the edge of the SNR
suggests that the FUV profiles are primarily broadened by thermal motions, i.e.,
that the emission is dominated by the limb-brightened edges where the 
bulk motion along the line of sight is small.  Since
the LWRS slit also covers most of \onine, we can expect the broad Ly $\beta$
profile from this SNR to also be shaped by thermal broadening.  Finally, we can
expect our MDRS spectrum of the \dem71\, blast wave to also exhibit lines 
dominated by thermal broadening, since the edge of the blast wave was targeted 
in this observation.  In all these cases, we have used the FUV line widths as
a direct measure of the ion temperatures and have utilized the non-radiative
shock code of Ghavamian \etal\, (2001, 2002) to predict the emission line
fluxes.

In contrast to the spectra mentioned above, the interpretation of our LWRS 
line profiles for \dem71\, is less straightforward.  In this case most of the FUV emission
is sampled from the projected center of the SNR, where emission from a range of viewing angles
(and hence bulk velocities) is sampled in the \fuse\, spectrum.
Furthermore, the fraction of the \dem71\, interior
covered by the LWRS slit is large enough to make a simple plane-parallel
approximation inadequate for describing the two sides                     
of the shell.  In the next section we describe a more realistic model
where we approximate the SNR as an expanding spherical shell.  
Having used these models to disentangle the bulk and thermal
contributions to the line broadening, we then compute the proper shock speed(s)
and $T_{p}$/$T_{O}$ ratio(s) required to match the FUV line profiles for
\dem71.  Finally, we model the 
flux ratios for \dem71\, using our numerical shock code and constrain
the electron-ion equilibration and the preshock neutral fraction.

\subsection{NUMERICAL MODELS}

We begin by describing our plane parallel shock models, then describe
the spherical shell model for \dem71.  We use the former models to predict
flux ratios for the broad Ly $\beta$ and O~VI emission lines, and use the
latter models to match the shapes of these lines in the LWRS spectrum of
\dem71.

Before presenting our models, we note that Heng \& McCray (2006; hereafter HM06)
have recently presented new calculations of Balmer-dominated shock structures.
In their models HM06 found that the fast neutral distribution evolves
in shape over multiple charge exchanges, with the overall broad component
profile being the sum of multiple profiles (this effect was ignored
in our earlier works).   They find that for \vs$\,\gtrsim$\,2000 \kms\,
the fast neutrals become ionized before they have undergone enough
charge exchanges to fully sample the bulk-shifted velocity distribution of the incident 
hot protons.  The resulting fast neutral distribution is shifted by a
bulk velocity substantially less than $\frac{3}{4}$\,\vs, and is narrower
than predicted from the models of Chevalier, Kirshner \& Raymond (1980) and our
own models (Ghavamian \etal\, 2001, 2002).  HM06 find 
this effect to be far less important at the shock speeds relevant to \dem71, so
the shell model results presented for that SNR in Section 5.1.1 should remain valid.
However, in the cases of \onine\, and \onenine\, where shock speeds are expected 
to exceed 2000 \kms, our interpretation of the FUV line profiles changes significantly 
if we adopt the HM06 mapping between shock speed and broad
hydrogen line width.  In the HM06 calculations the range of shock speeds
implied by a given broad H~I line width is significantly narrower in the
limits of minimal to full electron-proton equilibration (for \vs\,$\gtrsim$2000 \kms)
than in our line profile calculations (Ghavamian
\etal\, 2001).  We describe these results in more detail in Sections 5.2 and 5.3.

\subsubsection{SHOCK MODEL}

The numerical code described by Ghavamian \etal\, (2001; 2002) computes
the ionization and temperature structure behind a non-radiative shock with
arbitrary electron-ion and ion-ion equilibration.  The original purpose of the
code was to compute the broad-to-narrow flux ratio of the Balmer H$\alpha$ and
H$\beta$ lines, including the enhancement of the narrow Balmer lines
by Lyman line radiative transfer.  We have now extended the ionization structure
calculation of this code to include heavier elements up to and including silicon.  In addition
to the electrons and protons, our code now includes Coulomb equilibration 
of the He, C and O ion temperatures behind the shock.   

At the large shock velocities expected in \onenine\,
and \onine\, ($V_{S}\,\gtrsim$4000 \kms), collisional ionization of neutral
hydrogen by He$^{+}$ ions and alpha particles becomes important (Laming \etal\, 1996), so we have included
these processes in the code via rates computed with the atomic close-coupling cross sections 
of Toshima (1994).  We have also updated the ionization balance to
include charge exchange between neutral hydrogen
and both He$^{+}$ ions and alpha particles, again utilizing the cross sections
computed by Toshima (1994).  All our rate calculations assume Maxwellian particle distributions
and are computed in the center of mass frame of the incident and target particles in
the manner described by Ghavamian (1999).
The other update included in this shock code is the calculation of emission line
fluxes in broad Ly $\gamma$, C~III $\lambda$977, broad Ly $\beta$ and O~VI 
$\lambda\lambda$1032, 1038 out the front of the shock.  We computed the collisional excitation of these
lines by electron, proton, He$^{+}$ and alpha particle impact.  The atomic
data used in our calculation utilize fits to collision strengths for O~VI
excitation by electron impact from the compilation of Merts \etal\, (1980).
To compute heavy ion excitation rates we have integrated the proton and alpha
particle excitation cross sections of O~VI from Laming \etal\, (1996), again under
the assumption of Maxwellian particle distributions.
We have also computed Ly $\beta$ and Ly $\gamma$ rates
from He$^{+}$ and alpha particle impact from the scaled cross sections of
Janev (1996). 

Although all of the observed emission lines arise from resonant
transitions, we have ignored the radiative transfer of the FUV photons
within the shock in our calculations.  The integrated resonance scattering optical depth perpendicular
to the shock front in the broad Ly $\beta$, Ly $\gamma$, C~III and O~VI lines is generally
small, $\lesssim$0.03 for a preshock density of 1 cm$^{-3}$, preshock hydrogen neutral 
fraction of 0.5 and shock speeds in the range 600$-$5000 \kms.  In addition, the very
large Doppler widths of the FUV lines ($\sim$600$-$3000 \kms\, for the SNRs considered
here) ensure that most photons propagating perpendicular to the shock front escape
before being absorbed.  The prominent absorption features observed at the centers of
the emission lines are almost entirely due to LMC and Galactic halo absorption.

Before describing the spherical shell model for \dem71\, we consider the question of
whether the ion-ion equilibration can be measured separately from the electron-proton
equilibration in the LMC Balmer-dominated SNRs.
At shock speeds greater than 3000 \kms, FUV line emission from collisional excitation
by protons, He$^{+}$ ions and alpha particles becomes increasingly important.  This implies that
the flux ratios can be sensitive not just to the electron-proton equilibration, but the proton-helium
ion equilibration as well. Since there are no He lines in our \fuse\, spectra of the Balmer-dominated SNRs, 
we have no a priori knowledge of the degree of proton-He ion equilibration in these objects.
This raises the question of whether flux ratios uniquely determine the
electron-proton equilibration.   The \lybovi\, ratio is most sensitive
to $T_{He}/T_{p}$ in shocks with minimal electron-proton equilibration.  For
\tetp\,$\lesssim$\,0.05, the ratio \lybovi\,
for shocks in the range 3500$-$8000 \kms\, (appropriate for \onine\, and \onenine) increases 
by only 10\% as $T_{He}/T_{p}$ is varied from 4 (i.e., $m_{He}/m_{p}$) to 1.  Therefore, while 
the FUV line fluxes are sensitive to \vs\, and \tetp\, in \onine\, and \onenine\, (\vs\,$\gtrsim$3000 \kms),
they do not vary strongly with ion-ion equilibration over the range of shock speeds 
set by the width of the broad Ly $\beta$ profile.

\subsubsection{SPHERICAL SHELL MODEL}

To predict the shapes of the FUV line profiles in our \fuse\, spectrum of \dem71\, we have
computed the emergent intensity in broad Ly $\beta$ and O~VI from two limb-brightened,
hemispherical shells joined together in the plane of the sky.   For each shell 
the shock speed, preshock density and degree of temperature equilibration between
the protons and oxygen ions are free parameters.   From Figure~1 we estimate that
approximately half the \dem71\, shell is subtended by the LWRS slit.  

The first step in computing the line profile within the aperture is to compute the emission 
at a velocity $v$ and an angle $\theta$ from the center of a given shell via the well known radiative
transfer relation

\begin{equation}
I_{v}(\theta)\,\,=\,\,S\,(1\,-\,e^{-\tau_{v}(\theta)})
\end{equation}

\noindent where $S$ is the source function of the shell and $\tau$ is 
the optical depth of the shell at the azimuthal angle $\theta$.  Under the assumption of a 
Gaussian line profile, the optical depth for the front or back side of the shell is

\begin{equation}
\tau_v(\theta)\,=\,\,\tau_0\,\,exp\,(-(V\,\mp\,V_B\,cos\,\theta)^2/b^2)\,sec\,\,\theta
\end{equation}

{\noindent where} \vb\,$\equiv$\,$\frac{3}{4}$\,\vs is the bulk velocity of the postshock gas,
and $b\,\equiv\,(2 k T / M)^{1/2}$ is the Doppler parameter of the emission line.  
The $\mp$ symbol is negative for the approaching shell and positive
for the receding shell.  The face-on optical depth $\tau_0$ is evaluated at the center of the line
profile,

\begin{equation}
\tau_0\,\equiv\,\frac{\sqrt{\pi} e^2}{m_e c}\,\frac{f_{ij} \lambda_{ij}}{b}\,N
\end{equation}

\noindent where $N$ is the column density through the shock of the ion and
$f_{ij}$ are the oscillator strength and $\lambda_{ij}$ the rest wavelength
of the ionic resonance line.  In terms of these quantities, the source function $S$
is

\begin{equation}
S\,=\,\frac{I_0}{\sqrt{\pi} \tau_0\,b}
\end{equation}

\noindent In the limit $\tau_{0}\,sec\,\theta\,\ll\,$1, Equation (1) reduces
to the usual expression for limb-brightened emission $I = I_{0}\,sec\,\theta$,
where $I_{0}$ is the face-on emission line intensity.
If the shock speed, preshock density and postshock equilibration
are allowed to differ between the approaching and receding sides of the shell then
$S$, \vb, $b$ and $\tau_{0}$ will also differ between the two sides.
Although photons from the far side of the shell must traverse the
near side of the shell as they exit the SNR, we assume that there is negligible absorption
of these photons on the near side due to large velocity offsets.  Despite the moderate blending of the Ly $\beta$
and O~VI lines in the LWRS spectrum of \dem71, the optical depth for absorption of highly
Doppler shifted photons from one profile into another is negligible within the shock,
so we neglect this effect in our models.

The flux per velocity interval from each shell $F(v)$
is obtained by inserting Equation (2) into Equation (1) for each of the
two sides, then performing the integral 

\begin{equation}
F(v) \,\, \propto \,\, \int_{0}^{\Theta}\,I_{v}(\theta)\,cos\,\theta\,\,d\,(cos\,\theta)
\end{equation}

\noindent for a sample of velocities over the line profile (we assume azimuthal
symmetry).   The total observed profile is then the summed fluxes from the front and back sides 
of the shell.

In a spherical shell of radius $R$ and thickness $d$, the maximum limb brightening
is $sec\,\,\theta_{max}\,\approx\,\sqrt{R / 2 d}$, where $R$ is the radius
of the SNR and $d$ the thickness of the emitting layer.  The larger the radius
and the thinner the emitting layer, the greater the limb brightening.  
In our models we relate the upper limit on the angular integral in Equation~5,
$\Theta$, to the limb brightening factor via $\Theta\,=\,\alpha\,\theta_{max}$,
where 0$\,\leq\,\alpha\,\leq\,$1 is determined by the fraction of the SNR shell
covered by the spectrograph slit.  For a slit covering the full SNR, 
$\alpha\,=\,$1.  For a preshock density of 1 cm$^{-3}$, for example,
$d_{O~VI}\,\sim\,$(3-60)$\times$10$^{-3}$ pc between shock speeds of
500 and 5000 \kms.  Taking $R\,\approx\,$10 pc for \dem71, the limb brightening
factor lies in the range 9\,\,$\lesssim$\,$sec\,\,\theta_{max}\,\,\lesssim$\,40,
so that the integrated emission line profile $F(v)$ is dominated by 
the contribution near the edge of the shell.   

To predict the line profile shapes we have used our
numerical shock code to compute a grid of values for $\tau_{0}$ over a 
wide range of shock speeds and equilibrations.  
When computing the line profile for a given set of shock parameters,
we interpolated within the pre-calculated grid to compute the appropriate
value of $\tau_{0}$.  For a given shock speed between 500 \kms\, and 5000 \kms\,
the O~VI column density behind the shock is relatively insensitive to the
shock speed, but declines with increasing electron-proton equilibration,
dropping by a factor of 3 between the limits of minimal and full electron-proton
equilibration.  

For the range of shock parameters quoted above, the line center optical depths
of broad Ly $\beta$, O~VI $\lambda$1032 and C~III $\lambda$977 in our models can reach values
$\sim$0.5-1 at the edge of the SNR shell.  At these moderate optical depths the resonance
scattering removes photons from the line center and leaves emission line profiles
with slightly flattened, broadened peaks.  In the case of the O~VI doublet the
$F(\lambda1032)$:$F(\lambda1038)$ ratio drops below the optically thin ratio of 2:1,
an effect that has already been observed in non-radiative shocks in the
Cygnus Loop (Long \etal\, 1992; Sankrit \& Blair 2002;  Raymond \etal\, 2003).
Although consideration of resonance scattering is
important for interpreting the spectra of a localized filament in
a nearby object such as the Cygnus Loop, the net effect on the spectra of the
LMC Balmer-dominated SNRs is rather small.  In cases where the \fuse\, aperture
includes edge-on emission from the SNR,
enough O~VI $\lambda$1032 photons are contributed from
positions just inside the edge of the shell to compensate for photons scattered
out of the line of sight from positions at the edge of the shell.  Our simple
shell models do not include these effects, so they may over-estimate the
importance of resonance scattering on the FUV profiles of the LMC remnants. 
In comparing line fluxes from these objects with predictions from our plane parallel
shock models we have neglected the effects of resonance scattering.

The line profiles calculated with the spherical
shell model are flatter on top and exhibit slightly more extended wings than
Gaussian profiles.  However, given the strong interstellar absorption near the centers of the
FUV lines and the limited signal-to-noise of most FUV line profiles, both
the single Gaussian profile and the spherical shell model adequately match
the line shapes in the \onenine\, and \onine\, spectra.

\section{MODEL RESULTS}

In previous modeling of Balmer-dominated SNRs (Ghavamian \etal\, 2001; 2002),
we estimated the degree of electron-proton temperature equilibration by
first measuring the width of the broad H$\alpha$ component from edge-on
shocks to constrain the range of shock speeds, then modeling the broad to narrow flux ratio
to constrain the combination of shock speed and \tetp\, consistent with all the data.
Unfortunately, the narrow Ly $\beta$ emission is completely absorbed by the ISM in our \fuse\,
spectra, as is the core of the broad Ly $\beta$ line.  In addition, the widths of the FUV emission 
lines in all our \fuse\, spectra reflect both thermal and bulk Doppler broadening.
Without taking into account the latter contribution to the line profile, the shock
velocities tend to be overpredicted.

In the ensuing analysis we take the following approach to constraining the shock kinematics:
(1) We use existing measurements of the broad component H$\alpha$ width (where available) from
the rim of each SNR to constrain the range of shock speeds.  The line widths
from these measurements should primarily reflect thermal broadening and
give the best constraint on the appropriate range of \vs.  (2) Using the range
of constrained shock speeds from the plane parallel models, we calculate
fluxes in broad Ly $\beta$ and O~VI.
Comparing these model fluxes with the observed values, we attempt to further refine
our estimates of \vs\, and \tetp, while also obtaining constraints on the preshock neutral fraction of hydrogen.
(3) Again utilizing the shock speed constraints from the plane parallel models, we 
use our spherical shell models to predict the shapes of the integrated Ly $\beta$ 
(and when possible, O~VI) profiles.  Comparing these predictions 
with observed profiles in the LWRS spectra, we gauge the contribution
of bulk Doppler broadening to the profile width.  By subtracting the
bulk Doppler component in quadrature from the total profile width
in Table~2, we obtain limits on $T_p$ and $T_O$, thereby constraining \totp.

In the following we compare the observed \lybovi\, with predictions from our plane parallel shock 
models to constrain \tetp\, for the \dem71\, blast wave, \onine\, and \onenine.   In all calculations
we adopt the LMC abundances of Russell \& Dopita (1992)
(in this case by 12 + log(O/H) = 8.35).

\subsection{\dem71}

\subsubsection{SOUTH RIM OF \dem71}

The FWHM of the broad Ly $\beta$ line in our MDRS spectrum is 1365$\pm$75 \kms, nearly 40\% larger than the
H$\alpha$ width of 805$^{+140}_{-115}$ \kms\, measured by Ghavamian \etal\, (2003) from 
Rutgers Fabry-Perot (RFP) observations of \dem71.  There are several potential explanations for this
discrepancy.  First, there was a lack of extensive spectral
coverage on either side of the H$\alpha$ profiles in the RFP spectra of Ghavamian
\etal\, (2003).  Since the H$\alpha$ profiles were not sampled beyond $\pm$800 \kms,
fitting a line with wings extending well beyond that range may have resulted in an
anomalously low line width.  On the other hand it is possible that contrary to our
expectation, the MDRS observation of \dem71\, did not completely isolate the limb emission.
In recent optical spectra of \dem71\, acquired with the Magellan 6.5m telescope (C. Rakowski
2006; private communication) several broad H$\alpha$ components are observed along portions of the
eastern rim.  The two components are of comparable width ($\sim$600 \kms), but their
line centroids are shifted in opposite directions in velocity.  This indicates that even
near the limb of the SNR the slightly off-edge
emission can be bright enough to significantly affect the
observed broad component profile.  The multiple nested filaments observed at the southern
limb of \dem71\, (Figure~2) suggest multiple tangencies between the blast wave and line of sight,
indicating that the emission between these tangencies is likely to include partially face-on
emission.  Unlike the RFP spectral extraction, which cleanly isolated the H$\alpha$ emission from the southern
rim of \dem71, the MDRS slit is wide enough to include some of this off-tangent emission, so the
MDRS spectrum is more likely to be affected by this problem. If this explanation is correct, then the line widths
measured in our MDRS spectra would overpredict the shock velocity.  

With the above caveats in mind, we modeled 
the \dem71\, MDRS spectrum by first using the H$\alpha$ line width of 805$^{+140}_{-115}$
\kms\, from Ghavamian \etal\, (2003) to set the range of shock speeds.  Since the
shock speed is likely to remain relatively constant near the rim of \dem71, we
assume that the FUV line ratios reflect excitation from a single shock, even though
the lines exhibit some bulk broadening.  Under these assumptions, we then compared
the predicted \lybovi\, ratios with the observed values.  

Using the relationship between hydrogen line FWHM and shock speed
calculated by Ghavamian \etal\, (2001), the implied range of shock speeds from the
H$\alpha$ broad component width (Ghavamian \etal\, 2003) is 775$-$1005 \kms\,
over the range \tetp\,=\,$m_{e}/m_{p}$ to 1.  We calculated a grid of \lybovi\, values
over this range of shock speeds, combining each \tetp\, with the corresponding velocity needed
to match the width of the broad Ly $\beta$ line.  A plot of our results is shown in the top panel of
Figure~8.   

For the range of shock speeds here, the electrons in the H~I neutral zone behind the shock are close to the
optimum temperature ($\sim$10$^{6}$~K) for exciting Ly $\beta$ close to the shock front, producing a peak in the
\lybovi\, curve near \tetp\,=\,0.  At higher equilibrations and shock speeds, however, the charge exchange
rate begins to dominate over the collisional ionization rate, resulting in higher overall density of
fast neutral hydrogen behind the shock.  The broad Ly $\beta$ flux increases accordingly, producing
the rise in \lybovi\, seen in Figure~8 at high equilibrations.

Unfortunately, due to the linear dependence of \lybovi\, on
preshock neutral fraction, the double valued behavior of the predicted \lybovi\, at
low and high \tetp, and the uncertainty in the measured flux ratio,
we are unable to obtain a unique combination of \tetp\, and \vs\, from the plot.  Although we cannot
unambiguously determine these two quantities, we can at least constrain the range of
preshock hydrogen neutral fractions in the southern rim of \dem71.  If we require
at least one point of intersection between the range of measured and predicted \lybovi,
the maximum and minimum allowed preshock neutral fractions are 40\% and 20\%, respectively.

Assuming a uniform shock speed along the southern rim of \dem71, 
we have attempted to estimate the degree of proton-oxygen equilibration at the shock transition.
At the range of shock speeds considered here (800$-$1000 \kms), charge exchange behind the shock produces
a fast neutral distribution with a velocity width nearly equal to the width of the proton distribution.
Therefore, the proton temperature is just the temperature obtained from the broad
component H$\alpha$ width (805 \kms).  If we assume that the difference between this value
and our measured Ly $\beta$ width is due to the effect of bulk motion of emitting material
within the \fuse\, aperture, then we can estimate the Doppler broadening as
$\sim$ (1365$^2$ - 805$^2$)$^{1/2}$ \, = \, 1100 \kms.  This value, however, is larger than
the measured O~VI line width of 935 \kms.  Therefore, $T_O$ is not well constrained
from our fit to the O~VI profile, and we can only estimate an upper limit on \totp under the assumption
that the O~VI width represents pure thermal broadening.  Under these assumptions,
\totp\,=\,$\frac{m_{O}}{m_{p}}\,\frac{\sigma_{O~VI}^{2}}{\sigma_{p}^{2}}$, giving
\totp\,$\lesssim$\,8.  Note that the oxygen temperature derived from the
O~VI line width represents an average value weighted by the postshock O~VI emissivity.
The O~VI zone behind the shock is extended enough for Coulomb collisions to
have transferred some energy from the oxygen ions to the lighter ions.  Comparing our measured
values of \totp\, with predictions from shock models, we find the emission-weighted
ratio should be $\sim$10\% smaller than the immediate postshock value, so that 
\totp\,$\lesssim\,$9 is a better estimate of the temperature
ratio.  Clearly this is a rough limit, but it suggests that
a moderate amount of proton-oxygen temperature equilibration has
taken place at the shock transition.  

\subsubsection{LWRS SPECTRUM OF \dem71}

As described earlier, the LWRS slit in the \dem71\, observation is filled mostly
with emission from the interior of \dem71.  Therefore,
we expect each of the emission lines in Figure~3 to consist of two
components, one red-shifted and one blue-shifted from the rest frame of the SNR.  Utilizing
our spherical shell model for \dem71\, we have attempted to match the shapes of the broad Ly $\beta$
and O~VI $\lambda$1032 lines.  The faintness of the C~III $\lambda$977 feature prevents us from obtaining
meaningful constraints on its shape, so we have not attempted to model the C~III $\lambda$977 profile.

We focused first on separating the thermal and geometric contributions to the width
of the broad Ly $\beta$ line.  We attempted to match the profile shape with our spherical shell models 
by varying \vs\, and \tetp\, together for each of the two sides of the shell contained in the aperture.
The amplitudes of the frontside and backside profiles are not allowed to vary independently, but are
tied together via the requirement of ram pressure conservation: $n_{b}\,v_{b}^{2}\,=\,n_{f}\,v_{f}^{2}$.
We do not model the LMC and Galactic Halo absorption 
in these models, but focus upon matching the wings of the observed Ly $\beta$ profile.

Our best estimate of the Ly $\beta$ profile shape from the shell models is shown in
Figure~9.  As a first estimate, we assumed
the LWRS slit was aimed at the center of the SNR
and covered half the projected area of that shell (approximately correct for our observational
setup).  Interestingly, the line profiles predicted by this model (dashed line in Figure~9) overpredict 
the emission near the center of the
observed profile, and give only slightly better agreement in the far wings of the profile.  Varying
the amplitude or width of the model profile for
any individual component (approaching or receding side of the shell) can produce better agreement
toward the center of the broad Ly $\beta$ line, but the agreement then worsens in the wings of the profile.
Since the profiles from the front and back sides of the shell
are centered on the bulk velocities (=$\frac{3}{4}$\vs) of these two components, the blueshifted
and redshifted peaks are spread far apart, with too
little emission predicted at the center of the profile.  We found that the
only solution to this problem was to include emission from edge-on shocks in the profile calculation
(i.e., by setting $\alpha\,=\,$1 in Equation~5),
despite the fact that the LWRS observation does not appear to include filaments near the edge
of \dem71\, (Figure~1).  Under that assumption the wings of the calculated profile draw inward, with emission from
zero radial velocity filling in the center of the broad Ly $\beta$ line.  
The necessary inclusion of an edge-on component
in our models indicates that some of the O~VI emission may be distributed differently
from the H$\alpha$ emission (particularly if parts of the \dem71\, blast wave are propagating
into fully ionized gas).

While the addition of the edge-on component significantly improves the ability of our models to match the observed profile,
asymmetric flux contributions from the front and back sides of the shell are also required to match
the residual redshift of the broad Ly $\beta$ centroid.  To reproduce the +235 \kms\, redshift of the
broad Ly $\beta$ feature from the LMC rest velocity (Table~2), the central peak of the Ly $\beta$ profile
from the far side of the shell must be higher that of the near side.
Keeping this constraint in mind, we adjusted the shock velocity, total preshock density
and electron-proton equilibration for the two sides of the shell while conserving
ram pressure as described earlier.  The final
result of our Ly $\beta$ profile calculation for \dem71\, is marked
with the solid line in Figure~9.

Using the calculated relationship between the velocity width of the proton distribution and
the width of the H~I profile transmitted by charge exchange (Chevalier, Kirshner \& Raymond 1980;
Ghavamian \etal\, 2001), the best combination of Ly $\beta$ profile 
widths from our spherical shell model are 920 \kms\, for the approaching side of the shell and
585 \kms\, FWHM for the receding side of the shell.  The corresponding combination of
shock parameters (\vs, $n_{H^{0}}$, \tetp) are (900, 0.5, 0.05) for the
approaching side of the shell and (650, 1.0, 0.6) for the receding side of the shell.  Note
that the absolute quoted values of $n_{H^{0}}$ are not the meaningful quantities here, but rather
their ratio, as required by ram pressure conservation.

Although there is some degeneracy between the various shock parameters estimated above, 
they are fully consistent with prior measurements of the broad component
H$\alpha$ width, shock speed and equilibration estimated for \dem71\, by Smith \etal\, (1991), Ghavamian \etal\, (2003) and Rakowski
\etal\, (2003).  The analyses in these papers indicated that the blast wave of \dem71\,
has decelerated to $\sim$600 \kms\, in the northeastern corner of the SNR, precisely where the
\fuse\, LWRS slit grazes the edge.  On the other hand, the RFP H$\alpha$ spectra and \chan\, X-ray analyses
of Ghavamian \etal\, (2003) and Rakowski \etal\, (2003) indicated a nearly 50\% variation
in shock speed around the rim of of \dem71, with the highest shock speeds ($\sim$800-1000 \kms)
measured along the southern rim.  The results of our broad Ly $\beta$ profile models
indicate that the far side of \dem71\, may be propagating into a higher density medium
than the near side, with a correspondingly lower shock speed on the far side of the SNR.

After applying our shell models to the broad Ly $\beta$ profile, we modeled the O~VI $\lambda$1032
profile.  As a first guess we set the shock speeds of the front and
back sides of the shell equal to the values derived from the broad Ly $\beta$ profile model, with
the O~VI line width a free parameter.   Our best estimates for the O~VI line widths are 
475 \kms\, FWHM for the receding shell and 900 \kms\, FWHM for the approaching shell.  The centroid of the
observed O~VI $\lambda$1032 feature is shifted by the same velocity as the centroid of the Ly $\beta$
feature (Table~2), and can be matched with a 2:1 ratio of preshock density between the back and
front sides of the shell.   Taken together, the Ly $\beta$ and
O~VI $\lambda$1032 results suggest that while the front and back sides of \dem71\, are
propagating into an ISM with a 2:1 ratio in total density, the hydrogen neutral fractions
are similar on both sides of the shell.

From our above analysis, we can estimate \totp\, for the approaching and receding sides 
of the shell in \dem71.  Given \vs\,=\,650 \kms, \tetp\,=\,0.6 and an estimated O~VI
$\lambda$1032 width of 475 \kms\, on the back side, we obtain \totp\,$\approx\,$6.5.
On the near side, for shock speed of 900 \kms, \tetp\,=\,0.05, we obtain \totp\,$\approx$\,16.
Although these estimates are subject to the uncertainty of deconvolving the thermal
broadening from the bulk Doppler broadening in our spectra, both the magnitude and
declining trend of \totp\, with shock speed are consistent with results
established for the Balmer-dominated shocks in other SNRs such as the Cygnus Loop
(Raymond \etal\, 2003) and SN 1006 (Laming \etal\, 1996; Korreck \etal\, 2004).

\subsection{\onine}

The sole emission line positively detected in the LWRS spectrum of this Balmer-dominated
SNR is broad Ly $\beta$ (Figure~4).  The fitted FWHM (3710$\pm$400 \kms) is the 
largest hydrogen line width measured in a Balmer-dominated SNR to date.  
The centroid of the broad Ly $\beta$ line is substantially blueshifted ($-$540 \kms)
relative to LMC frame, suggesting that the FUV emission from \onine\, primarily arises
from the near side of the SNR shell.  It is likely that the broad Ly $\beta$ profile
is affected by bulk Doppler motions.  However, unlike the other three Balmer-dominated SNRs,
there has been no prior detection of broad ionic line emission from \onine, so we have no
clear constraint on the appropriate range of shock speeds for modeling the profile in Figure~4.

Since we do not have enough information for disentangling the bulk Doppler 
contribution of broad Ly $\beta$ from the thermal Doppler contribution, we
proceed with a plane parallel shock analysis under the crude assumption that the broad
Ly $\beta$ width in \onine\, reflects purely thermal broadening.  
Using the calculations of HM06, the measured FWHM corresponds 
to shock speeds of 5200$-$6300 \kms\, over the range $m_{e}/m_{p}\,\leq\,$\tetp$\,\leq\,$1
(for comparison, the FWHM calculations of Ghavamian \etal\, (2001, 2002) give 
a range of 5610$-$7780 \kms).  

Although the O~VI lines are not detected in the \fuse\, spectrum
of \onine, we can still place a lower limit on both \lybovi\,
and the widths of the O~VI lines.  First, we estimate an upper limit on the O~VI
$\lambda$1032 flux.  The upper limit on the flux per wavelength interval from the
\fuse\, spectrum of \onine\, is $\approx$ 2$\times$10$^{-15}$ ergs cm$^{-2}$ s$^{-1}$
\AA$^{-1}$ at the center of the O~VI $\lambda$1032 line.  To convert this number to
a flux, we require a limit on the O~VI line width.  For each shock speed in the range quoted 
earlier, the FWHM of the O~VI line is related to the oxygen temperature via
$V_{FWHM}(O~VI)\,=\,\,\sqrt{8\,ln\,2}\,\sqrt{k\,T_{O}/m_{O}}$, so that full
proton-oxygen equilibration would produce an O~VI line width $\frac{1}{4}$
as broad as the proton velocity distribution, while no equilibration would
yield equal widths.  Over the full range of allowed equilibrations,
the predicted O~VI line width ranges from 1245 \kms\, (for \tetp\,=\,\totp\,=\,1)
to 5300 \kms\, (for \tetp\,=\,1 and \totp\,=\,16).
\kms\, (equilibrated).  The corresponding limit on the O~VI $\lambda$1032 flux
is then F(1032)\,=\,(0.9-3.9)$\times$10$^{-14}$ \fl\, in the limits of full and minimal
proton-oxygen equilibration respectively.

In Figure~8 we present the predicted \lybovi\, ratio for \onine\, for the range
of shock speeds implied by the width of the broad Ly $\beta$ profile.  
The lower limit on the \lybovi\, ratio corresponds to the largest O~VI
line width ($\sim$5300 \kms) allowed from our analysis above.
For a given broad Ly $\beta$ width and corresponding combination of shock parameters (\tetp, \vs), 
the O~VI emission rises more rapidly with increasing \tetp\, (and shock speed) than
the broad Ly $\beta$ emission, causing the modest decline in the predicted \lybovi\,
(Figure~8).   Our models are shown for two limiting preshock neutral fractions:
$f_{H^{0}}\,=\,$0.4 and 0.8.  The lower neutral fraction is the smallest value allowing
an overlap between our model curves and the observed range.  The higher fraction
is the largest value likely for the warm ISM, given the minimum photoionization
expected from both the SNR and the ambient UV and soft X-ray backgrounds.

We can draw several conclusions from the middle panel of Figure~8.
First, efficient proton-oxygen ion equilibration is ruled out. In such a case,
the narrow O~VI line width (a minimum of 1245 \kms\, FWHM for \totp\,=\,1) would require
a very low total O~VI $\lambda$1032 flux to stay consistent with the non-detection
of O~VI in the LWRS spectrum.  This would raise the observed \lybovi\, ratio
(i.e., the lower edge of the shaded region) to  $\sim$ 9, which cannot be matched 
by any model regardless of equilibration or preshock neutral fraction.  The
prediction from our models is that \lybovi\,$\lesssim$4.5, a constraint which is
satisfied for $V_{FWHM}(O~VI)\,\gtrsim\,$2400 \kms.  This places the proton-oxygen
ion equilibration in the range 4\,$\lesssim\,T_{O}/T_{p}\,\leq\,$16 (the assumed upper limit corresponds to the
oxygen to proton mass ratio, though \totp\,$>$\,16 has been observed
on occasion in solar wind shocks).   We cannot place a strong constraint on \tetp\, 
from Figure~8, but we can deduce that the preshock gas should be significantly
neutral, 0.4\,$\lesssim\,f_{H^{0}}\,\lesssim$\,0.8.

We can use our derived shock velocity for \onine\, to estimate its age.
Assuming pure thermal broadening for the detected Ly $\beta$ line and 
taking into account the measurement uncertainty in the broad Ly $\beta$
width, the predicted shock speed is 3600$-$7100 \kms\, over the range
$m_e/m_p\,\leq$\,\tetp\,$\leq$\,1.   We compute the age $\tau$ by assuming
a radius of 3.6 pc (Warren \& Hughes 2004) and using $\tau\,\equiv\,\delta\,\frac{R}{V}$,
where $\delta$ is the expansion parameter.  Between the limits of free expansion and
the asymptotic Sedov-Taylor limit, $\delta$ ranges from around 1.0 down to 0.4.
Our derived shock speeds are significantly
lower than the 10,000-20,000 \kms\, expansion velocities predicted for Type Ia SNe,
indicating that some deceleration of the blast wave has occurred.  Assuming standard
Type Ia parameters and a constant ISM density, Dwarkadas \&
Chevalier (1998) predicted that $\delta$ declines to around 0.6 during the first few hundred
years after the SN explosion.  Using this expansion parameter, we find $\tau\,\approx\,$295-585
years for \onine.  This value compares favorably with the light echo age of 400$\pm$120
years found for SNR by Rest \etal\, 2005).
Given the neglect of bulk Doppler broadening in our analysis, we are at the very least encouraged
by the broad agreement between our derived age for \onine\, and the independent age estimate
of Rest \etal\, (2005).  

\subsection{\onenine}

\subsubsection{PLANE PARALLEL MODEL RESULTS}

The broad component H$\alpha$ width measured along the rim of \onenine\,
by Tuohy \etal\, (1982) was 2800$\pm$300 \kms, comparable to the value
measured in the Galactic Balmer-dominated SNR
SN 1006 (Ghavamian \etal\, 2002; Korreck \etal\, 2004).  From the
line profile calculations of HM06 this width
corresponds to shock speeds in the range 3200$-$3900 \kms\, in the
limits $m_{e}/m_{p}\,\leq\,$\tetp$\,\leq\,$1 (again these are
considerably narrower values than the range of 4115$-$5930 \kms\,
predicted by our own models).

The predicted \lybovi\, ratios for \onenine are shown in the bottom panel of Figure~8.
It is clear that in this case the flux ratios do not vary strongly with electron-proton
equilibration, partly a result of the narrow range of shock speeds implied by the
width of the broad Ly $\beta$ line.  The \lybovi\, curves are shown
for two limiting preshock neutral fractions: the lower fraction ($f_{H^{0}}\,=\,$0.4) being the smallest
value allowing overlap between our model curves and the observed range, with the
higher fraction ($f_{H^{0}}\,=\,$0.8) being the largest value likely for the warm ISM
surrounding the SNR. We cannot place a strong constraint on \tetp\, from
Figure~8. However, the preshock neutral fraction is well constrained,
0.4\,$\lesssim\,f_{H^{0}}\,\lesssim$0.5.  

Despite the considerable uncertainty ($\sim$40\%; Table~2) in the O~VI line width,
we can still place reasonable limits on the degree proton-oxygen ion equilibration in the
blast wave of \onenine.  The FWHM of the broad Ly $\beta$ line is
3130$\pm$155 \kms, while that of the O~VI $\lambda$1032 is 4975$\pm$1830 \kms.
The contribution to the FUV profiles from bulk Doppler broadening, $V_{B}$, can be estimated
by subtracting the Tuohy \etal\, (1982) H$\alpha$ width from the observed
Ly $\beta$ width, $V_{B}\,\approx\,\sqrt{3130^2 - 2800^2}$\,=\,1400 \kms.
Assuming the same bulk broadening for the O~VI line, the thermal width
of the oxygen ion distribution is then 4800$\pm$1750 \kms.  Since the O~VI line
width cannot significantly exceed that of broad Ly $\beta$, our result can only be
consistent with equal line widths, or \totp\,$\approx\,$16.

The age obtained by Rest \etal\, (2005) from their light echo study of \onenine\,\,
is 600$\pm$200 years.  Taking into account the measurement uncertainty in the broad
H$\alpha$ width, the full range of shock speeds for \onenine\, between the limits
of minimal and full electron-proton equilibration is 2600$-$4500 \kms.  Taking
$\delta\,\approx\,$0.6, this results in an age of 520$-$900 yrs, in good overall
agreement with the age predicted by the light echo study.

\subsubsection{SPHERICAL SHELL MODEL RESULTS}

Using the above constraints on the shock speed and equilibration, we
have computed the shape of the broad Ly $\beta$ line produced by a spherical shell.
Using Figure~7 of HM06 we find that for an H$\alpha$ line width
of 2800 \kms\, (the edge-on shock value of Tuohy \etal\, 1982) the predicted bulk velocity of
the fast neutrals is $\sim\,\frac{1}{2}$\,\vs, rather than the usually
assumed value of $\frac{3}{4}$\,\vs.  In that case
2000\,$\lesssim\,V_{B}\,\lesssim$\,3000 \kms\, for \onenine\, in the range
$m_e/m_p$\,$\leq$\,\tetp\,$\leq$\,1.  In addition, since the centroids of the FUV emission lines in \onenine\,
(Table~2) are consistent with the LMC rest velocity, we assume equal shock speeds and
preshock densities for the front and back hemispheres.  Using all of these parameters,
we were able to match the observed broad Ly $\beta$ profile with a thermal broadening
FWHM of 2500 \kms, consistent with the lower bound on the H$\alpha$
width of Tuohy \etal\, (1982).  The corresponding shock speed is approximately
2600 \kms\, from Figure~7 of HM06, with a correspondingly low
electron-proton equilibration (\tetp\,$\lesssim$0.2).  This result is consistent
with shock speeds at the lower end of the range predicted in the previous section.
It worsens the disagreement between the light echo age of Rest \etal\, (2005)
and our dynamical age estimate, and may provide further evidence of blast
wave deceleration.

\section{THE EFFECT OF ASSUMED ABUNDANCES}

It is clear that our interpretations above are significantly affected by our
choice of oxygen abundance via the linear dependence of the \lybovi\, ratio on the O/H abundance ratio.
Hughes, Hayashi \& Koyama (1998)
obtained a lower average oxygen abundance than Russell \& Dopita (1992)
from their analyses of {\it ASCA} spectra of LMC SNRs: 8.21$\pm$0.07 dex.  
It is instructive to consider the effect of such lower O/H ratios on the comparison of our shock
models with the observations.  If ISM oxygen abundances around all four Balmer-dominated SNRs were
set to the values derived by Hughes, Hayashi \& Koyama (1998), the modeled O~VI $\lambda$1032 
flux would decrease by 28\%, raising the predicted \lybovi
ratios in Figure~8 and lowering the range of acceptable preshock neutral fractions.  
In the case of \dem71, the predicted range of acceptable preshock neutral fractions 
becomes 15\%-30\%, while a fraction above 30\% is allowed for \onine.  The neutral
fraction for \onenine\, is most tightly constrained, with a value close to 30\% required
to match the new \lybovi ratio.

\section{REDDENING CONSTRAINTS FROM THE FUV SPECTRA OF \dem71\, AND \onenine }

Aside from the kinematic diagnostics available in the FUV spectra of the Balmer-dominated
SNRs, the ratio of Ly $\beta$ to Ly $\gamma$ flux is sensitive to the interstellar reddening,
providing further useful information on these objects.  Over the range of \vs\, and
\tetp\, considered in this paper, the impacting electrons and ions have kinetic energies
greatly exceeding the energy difference between the n\,=\,3 and n\,=\,4 levels of hydrogen,
and the ratio of these two lines reduces to the ratio of their collision strengths.  In
addition, the Ly $\beta$ and Ly $\gamma$ lines are optically thin due to their
large Doppler widths and the low density of the postshock gas.  This allows us
to ignore radiative transfer of Ly $\beta$ and Ly $\gamma$ photons when computing
their flux ratio.  The ratio is then nearly constant, ranging from approximately
3.3 for minimal equilibration to 3.1 for maximal equilibration.

In the LWRS spectra of \dem71\, and
\onenine, where emission from both lines has been detected, the Ly $\beta$ / Ly $\gamma$ ratios
are 3.7$\pm$0.4 and 3.4$\pm$0.5, respectively. Comparing these numbers with the predicted
values, these numbers are consistent with low to minimal reddening.  Koornneef \& Code (1981) noted that the slopes
of the Galactic and LMC extinction curves are very similar at $\lambda$\,$<$\,1400 \AA.  This implies
that for the ratio of two line fluxes in this wavelength range, the correction will be
nearly identical whether we use an LMC or Galactic extinction law.  Indeed, the
range of extinctions we derive for \dem71\, and \onenine\, using a Galactic
reddening law (Cardelli, Clayton \& Mathis 1989) are very similar to the
range derived for the non-30 Doradus regions of the LMC (Fitzpatrick 1985).
The color excess predicted for \dem71\,
is then E(B\,$-$\,V)\,=\,0.05-0.13, while for \onenine\, we find E(B\,$-$\,V)\,$\leq\,$0.11.

\section{DISCUSSION AND CONCLUDING REMARKS}

We have presented the first far ultraviolet spectra of the four known Balmer-dominated SNRs in 
the LMC.  These objects - \dem71, \onine, \onenine\, and \foureight, are the non-radiative (adiabatic) remnants 
of Type Ia SN explosions.  They feature optical spectra dominated by Balmer line collisional excitation of ambient
H~I gas overrun by the blast wave. Our \fuse\, observations resulted in the detection
of three of the four SNRs, \dem71, \onine\, and \onenine.   
The detected emission lines $-$ C~III $\lambda$977,
O~VI $\lambda\lambda$1032, 1038, Ly $\beta$ and Ly $\gamma$ $-$ are very broad (600 $-$ 3700 \kms) and are generated 
in the thin ionization
zones behind the non-radiative shocks.   In the case of \onine, the only detected feature
is faint, broad Ly $\beta$ (FWHM of 3710 \kms); however,
this is the first and only broad emission line detected in this SNR since its discovery by Tuohy \etal\, (1982).

The relative fluxes of the FUV emission lines are sensitive to the shock velocity, preshock neutral fraction
and the degree of electron-proton equilibration at the shock front.  On the other hand,
the widths of the FUV emission lines in Balmer-dominated SNRs are proportional to the relative
temperatures of the heavy ions, making them useful diagnostic probes of the degree of ion-ion temperature
equilibration in collisionless shocks.   These properties make FUV spectra of these SNRs
valuable tools for probing the collisionless heating in high Mach number
collisionless shocks. 

Comparing the \lybovi\, flux ratios
with predictions from plane parallel and spherical shell models of non-radiative shocks, we find that the observed
flux ratios are consistent to within the errors with low to minimal electron-proton equilibration
in shocks faster than 2500 \kms\, (observed in \onine\, and \onenine).  On the other hand, spectra
from the interior and limb of \dem71\, (\vs\,$\sim$600$-$1000 \kms) show evidence for small to moderate electron-proton
temperature equilibration.  Our results are generally
consistent with the inverse trend between electron-proton equilibration and Mach number derived
from optical (Ghavamian \etal\, 2001, 2002; Rakowski 2005, Ghavamian, Laming \& Rakowski 2007) and joint optical 
and X-ray (Ghavamian, Laming \& Rakowski 2007; Rakowski, Ghavamian \& Hughes 2003) observations of 
Balmer-dominated shocks in SNRs.

We have estimated \totp\, at the shock transition 
by comparing the broad Ly $\beta$ and O~VI emission line widths in our \fuse\, spectra of the
Balmer-dominated LMC remnants.  Over the range of shock speeds
sampled by our \fuse\, observations ($\sim$600\,$\lesssim$\,\vs\,$\lesssim$\,6000 \kms) our observations
indicate that \totp\, increases from $\sim$ 7 for shock speeds near 600 \kms\, to $\sim$ 16 (mass-proportional
heating) for shock speeds exceeding 1000 \kms.  A summary of our
modeling results is shown in Table~3.  From the shock velocities listed in Table~3
we obtain ages of 295-585 years for \onine\, and 520-900 years
for \onenine, respectively, in good agreement with the ages estimated for these SNRs
by SN light echo studies (Rest \etal\, 2005).

The main source
of uncertainty in our estimates of \totp\, is our uncertainty in the O~VI emission line widths.
In the case of \onenine, the faintness of the O~VI lines and their heavy blending with broad
Ly $\beta$ makes a precise measurement of their widths difficult.  In the case of \onine,
the O~VI emission falls below the detection threshold, allowing only an indirect estimate
of the O~VI line widths.  Although line blending is far less severe in our FUV spectrum of
the southern rim of \dem71\, all the emission lines are intrinsically very faint, again making
the measurement of the O~VI line widths (and hence \totp) uncertain.  Line blending is also
less severe in our \fuse\, spectrum of the interior of \dem71.  However, disentangling the frontside
from backside emission in this case is the dominant source of uncertainty in measuring \totp.

Modulo the bulk Doppler broadening, our results are consistent
with an inverse relationship between the degree of ion-ion equilibration and shock speed, with slower shocks 
exhibiting more prompt equilibration at the shock front than faster shocks.  This trend is similar
to that found for the electron-proton temperature equilibration, suggesting that the
collisionless processes responsible for
prompt electron-proton equilibration at low shock speeds may also cause prompt
ion-ion equilibration at low shock speeds.  A stronger verification of this trend will require
further measurements of \totp\, in other non-radiative shocks at intermediate velocities
(500$-$2500 \kms).  Obtaining these observations will remain the subject of future work.

P.G. would like to thank R. C. Smith for providing H$\alpha$ images of the
Balmer-dominated LMC remnants and J. M. Laming for helpful discussions on the atomic
physics of heavy ion excitation.  P. G. acknowledges support from NASA grants NNG04GL79G
and NAS5-32985, and W.P.B. acknowledges NASA grants NNG04GD150 and
NNG05GD75G, all to Johns Hopkins University.

\clearpage

\begin{deluxetable}{cccccc}
\tablecaption{List of FUSE Observations
                              \label{tblobs}}
\tablewidth{0pt}
\tablehead{
  \colhead{Object} & \colhead{Program ID} & \colhead{$\alpha_{J2000}$} &
  \colhead{$\delta_{J2000}$} & \colhead{Aperture} & \colhead{Exp. (ks)\tablenotemark{a}}
}
\startdata
SNR 0505$-$67.9 (DEM L 71)  &  P21403
     &  $05^{\rm{h}}05^{\rm{m}}42\fs7$ & $-$67\arcdeg\ 52\arcmin\ 38\farcs0
     &	LWRS	&  27.7 (12.7)  \\
SNR 0505$-$67.9 (DEM L 71 SW)  &  C072(03-04),C07206
     &  $05^{\rm{h}}05^{\rm{m}}39\fs7$ & $-$67\arcdeg\ 53\arcmin\ 14\farcs4
     &	MDRS	&  100.2 (44.6)  \\
SNR 0509$-$67.5 & P21402
     &  $05^{\rm{h}}09^{\rm{m}}31\fs9$ & $-$67\arcdeg\ 31\arcmin\ 17\farcs2
     &	LWRS	&  29.6 (9.7)   \\
SNR 0519$-$69.0 & P21401
     &  $05^{\rm{h}}19^{\rm{m}}33\fs8$ & $-$69\arcdeg\ 02\arcmin\ 09\farcs7
     &	LWRS	&  27.4 (12.7)  \\
SNR 0548$-$70.4  &  P21404
     &  $05^{\rm{h}}47^{\rm{m}}50\fs1$ & $-$70\arcdeg\ 24\arcmin\ 52\farcs1
     &	LWRS	&  29.5 (10.4)   \\
\enddata

\tablenotetext{a}{Total integration time for each object.  Orbital night time
fractions are listed in parentheses.}
\end{deluxetable}
 
\begin{deluxetable}{lccccc}
\tablewidth{0pt}
\label{tbl:table2}
\tablecaption{Measured Parameters for LMC Balmer-Dominated SNR Spectra\tablenotemark{a}}
\tablehead{
\colhead{Object} &
\colhead{Broad Ly $\beta$} &
\colhead{O~VI $\lambda1032$} &
\colhead{O~VI $\lambda1038$\tablenotemark{b}} &
\colhead{Broad Ly $\gamma$\tablenotemark{c}} &
\colhead{C~III $\lambda977$} 
}
\startdata
{\bf SNR 0505$-$67.9 (DEM L 71)}      &    &          &         &	&   \\
\,\,\,\,P214 (LWRS)	&    &          &         &     &   \\
~~~~~$\lambda_{cent}$ [\AA]  &  1027.1$\pm$0.05   & 1033.2$\pm$0.1  & 1038.9 (fixed) & 973.7 (fixed)  &  978.7 \\
~~~~~V$_{\rm cent}$ [km s$^{-1}$]         &  +410$\pm$15   &   +435$\pm$30    & +435 (fixed)  & +410 (fixed) & 480$\pm$50 \\
~~~~~V$_{\rm FWHM}$ [km s$^{-1}$] &  1135$\pm$30   &  740$\pm$45      & 740 (fixed)  &  1135 (fixed)   & 490$\pm$270 \\
~~~~~Flux [10$^{-14}$ ergs cm$^{-2}$ s$^{-1}$] & 37.2$\pm$1.9  &  6.0$\pm$0.4  & 3.0$\pm$0.3  & 10.0$\pm$0.8  & 1.2$\pm$0.3  \\
\,\,\,\,C072 (MDRS)  &    &          &         &     &   \\
~~~~~$\lambda_{cent}$ [\AA]  &  1026.8$\pm$0.1    & 1033.1$\pm$0.2  & 1038.8 (fixed) & \nodata  &  \nodata \\
~~~~~V$_{\rm cent}$ [km s$^{-1}$]         &  +320$\pm$30   &   +345$\pm$60    & +345 (fixed)  & \nodata & \nodata \\
~~~~~V$_{\rm FWHM}$ [km s$^{-1}$] &  1365$\pm$75   &  935$\pm$125      & 935 (fixed)  & \nodata   & \nodata \\
~~~~~Flux [10$^{-14}$ ergs cm$^{-2}$ s$^{-1}$] & 2.4$\pm$0.1  &  0.5$\pm$0.07  & 0.3$\pm$0.06  &  \nodata & \nodata  \\
  &    &          &         &     &   \\
{\bf SNR 0509$-$67.5}                 &    &          &         &   \\
\,\,\,\,P214 (LWRS) &    &          &         &     &   \\
~~~~~$\lambda_{cent}$ [\AA]  &  1024.8$\pm$0.24   & \nodata  & \nodata & \nodata  &  \nodata \\
~~~~~V$_{\rm cent}$ [km s$^{-1}$]         &  $-$263$\pm$125   &   \nodata   & \nodata    & \nodata  &  \nodata  \\
~~~~~V$_{\rm FWHM}$ [km s$^{-1}$] &  3710$\pm$400   &  \nodata   & \nodata   &  \nodata  &  \nodata \\
~~~~~Flux [10$^{-14}$ ergs cm$^{-2}$ s$^{-1}$] & 7.8$\pm$0.6  &  \nodata  & \nodata &  \nodata & \nodata  \\
  &    &          &         &     &   \\
{\bf SNR 0519$-$69.0}                 &    &          &         &   \\
\,\,\,\,P214 (LWRS) &    &          &         &     &   \\
~~~~~$\lambda_{cent}$ [\AA]  &  1026.8$\pm$0.24   & 1033.6$\pm$5.6  & 1039.3 (fixed) & 973.5 (fixed)  &  \nodata \\
~~~~~V$_{\rm cent}$ [km s$^{-1}$]         &  +320$\pm$70   &   +495$\pm$1630    & +495 (fixed)    & +320 (fixed)  &  \nodata  \\
~~~~~V$_{\rm FWHM}$ [km s$^{-1}$] &  3130$\pm$155   &  4975$\pm$1830      & 4975 (fixed)   &  3130 (fixed) &  \nodata \\
~~~~~Flux [10$^{-14}$ ergs cm$^{-2}$ s$^{-1}$] & 22.0$\pm$0.5  &  5.4$\pm$0.36  & 2.7 (fixed) & 6.4$\pm$0.9 & \nodata  \\
\tablenotetext{a}{Results quoted are for single Gaussian profile fits to each line}
\tablenotetext{b}{The velocity width and velocity centroid of the $\lambda$1038 line are 
tied to the width and centroid of $\lambda$1032 during spectral fitting. }
\tablenotetext{c}{The velocity width and velocity centroid of Ly $\gamma$ are fixed to those of Ly $\beta$}
\enddata

\end{deluxetable}

\begin{deluxetable}{lcccccccc}
\footnotesize
\tablewidth{0pt}
\label{tbl:table2}
\tablecaption{Shock Parameter Estimates for LMC Balmer-Dominated SNRs}
\tablehead{
\colhead{Object} &
\colhead{  } &
\colhead{Plane Parallel} &
\colhead{ } &
\colhead{ } &
\colhead{ } &
\colhead{Spherical Shell} &
\colhead{ } \\
\colhead{ } &
\colhead{V$_{\rm S}$ [km s$^{-1}$]} &
\colhead{\tetp} &
\colhead{\totp}  &
\colhead{ } &
\colhead{V$_{\rm S}$ [km s$^{-1}$]} &
\colhead{\tetp} &
\colhead{\totp}  
}
\startdata
{\bf SNR 0505$-$67.9 (DEM L 71)}      &    &          &     \\
~~~~~Blueshifted Shell (LWRS)  & \nodata \, & \nodata & \nodata & &  650   &   0.5    & 6.5  \\
~~~~~Redshifted Shell (LWRS)  & \nodata \, & \nodata & \nodata & & 900  &  0.05  &  16 \\
~~~~~Southern Edge (MDRS)  &  775-1005\tablenotemark{b}  &  $\leq$0.05\tablenotemark{b}  &  1-9 & & \nodata \, & \nodata & \nodata \\
  & & & &    &          &         &   \\
{\bf SNR 0509$-$67.5}                 &    &         &   \\
~~~~~Global Spectrum (LWRS)  &  5200-6300   &   \nodata  & 4-16 & & \nodata  &  \nodata  &  \nodata  \\
 & & & &    &          &         &   \\
{\bf SNR 0519$-$69.0}                 &    &          &   \\
~~~~~Global Spectrum (LWRS)  &  3200$-$3900   &   \nodata    & 16 & & 2600  &  $\lesssim$0.2  &  \nodata  \\
\tablenotetext{a}{When \tetp\, is unconstrained for plane parallel models, the shock velocities are quoted between the limits
of minimal and full electron-proton equilibration}
\tablenotetext{b}{Estimate from combined H$\alpha$ and X-ray spectroscopy of Rakowski, Ghavamian \& Hughes (2003) }
\enddata

\end{deluxetable}

\begin{figure}
\plotone{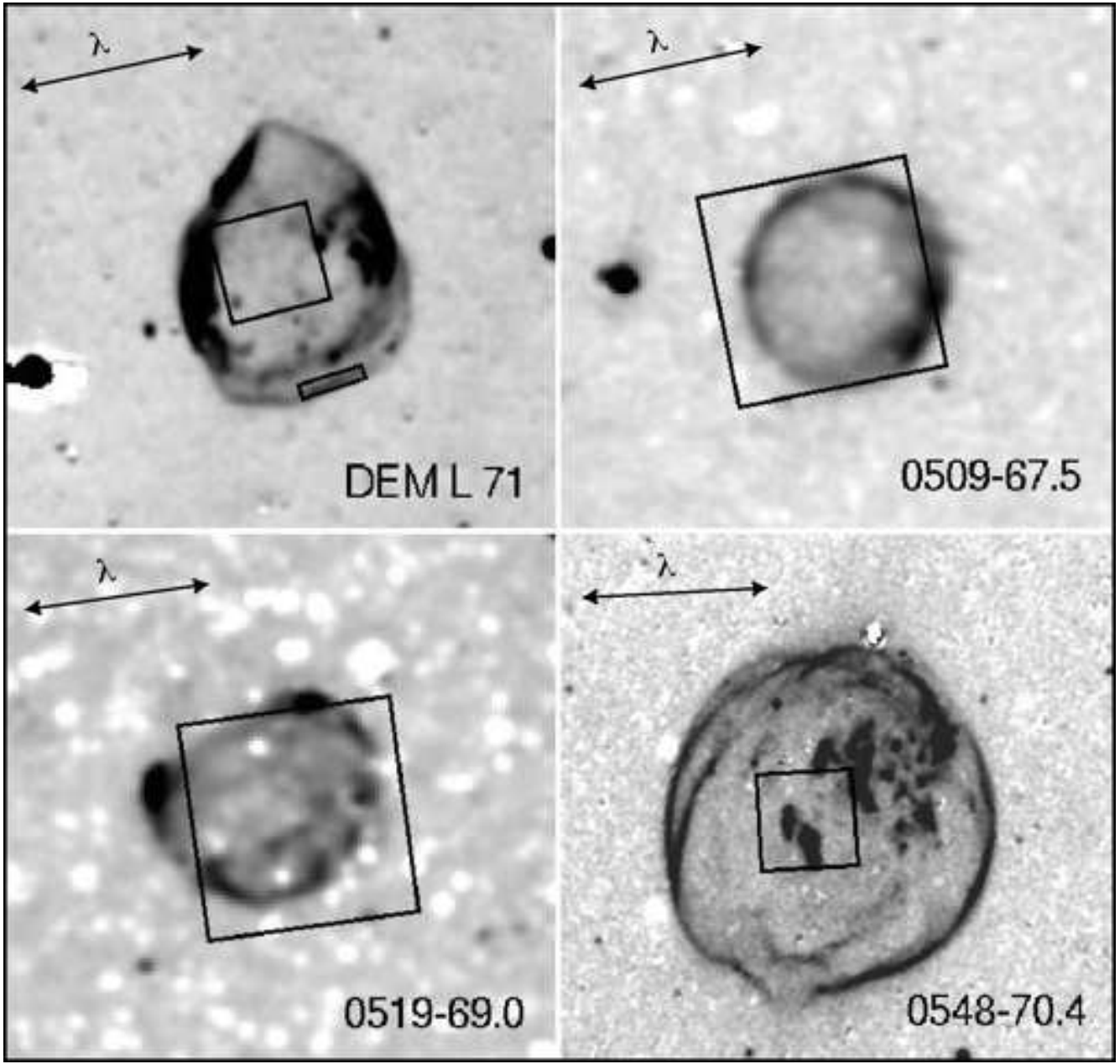}
\caption{Continuum-subtracted narrow band H$\alpha$ images of the four Balmer-dominated SNRs (CTIO 4m images
courtesy R. C. Smith).  The positions of the \fuse\, apertures (LWRS: 30\arcsec\,$\times$\,30\arcsec\, squares;
MDRS: 4\arcsec\,$\times$\,20\arcsec\, rectangle) are marked.
The wavelength dispersion direction for each of the LWRS spectra shown in Figure~2 are marked in the upper
left corner of each panel.   The dispersion direction for the MDRS pointing in \dem71\, is perpendicular
to the long dimension of the slit. }
\end{figure}

\begin{figure}
\plotone{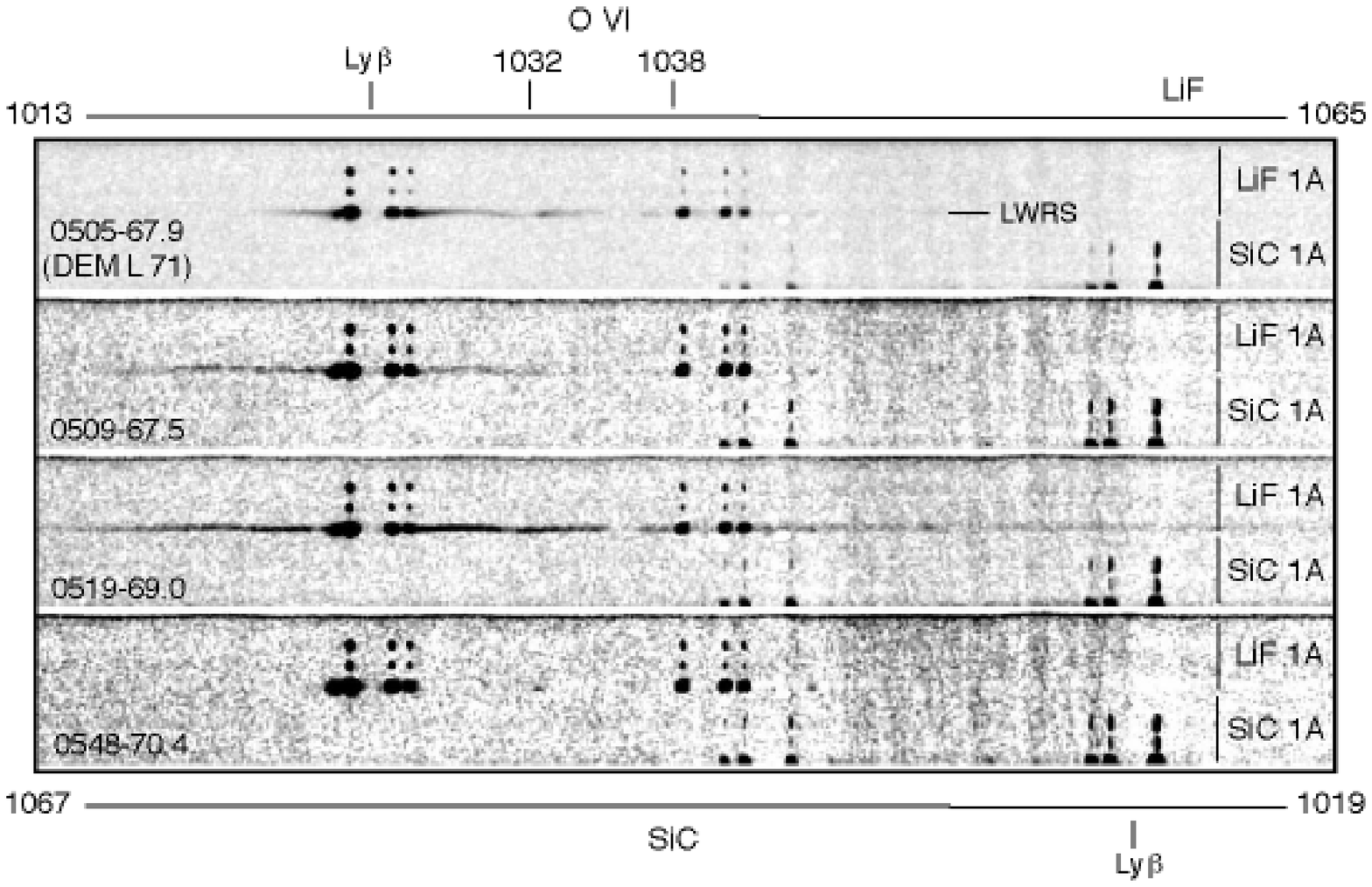}
\caption{Combined raw \fuse\, spectra of the four Balmer-dominated SNRs.  The 
1A channel data (1013 \AA\, $-$ 1067 \AA) are shown in the figure.  Positions of
major emission line features at the LMC systemic velocity (+275 \kms) are marked.  
The broad Ly $\beta$ and O~VI emission
lines are clearly seen in the spectra of \dem71, \onine\, and \onenine.  A particularly strong
C~II absorption feature can be seen in the spectrum of \onenine\, near 1037 \AA, arising from 
LMC halo absorption.  There
are two sets of spectra imaged on each detector (marked LiF 1A and SiC 1A).  Each
set consists of spectra from the medium resolution (MDRS, at top), high
resolution (HIRS, in middle) and low resolution (LWRS, bottom) apertures.  The
object spectra are acquired through the LWRS aperture, with the remaining
apertures containing only residual airglow.    }
\end{figure}

\begin{figure}
\plotone{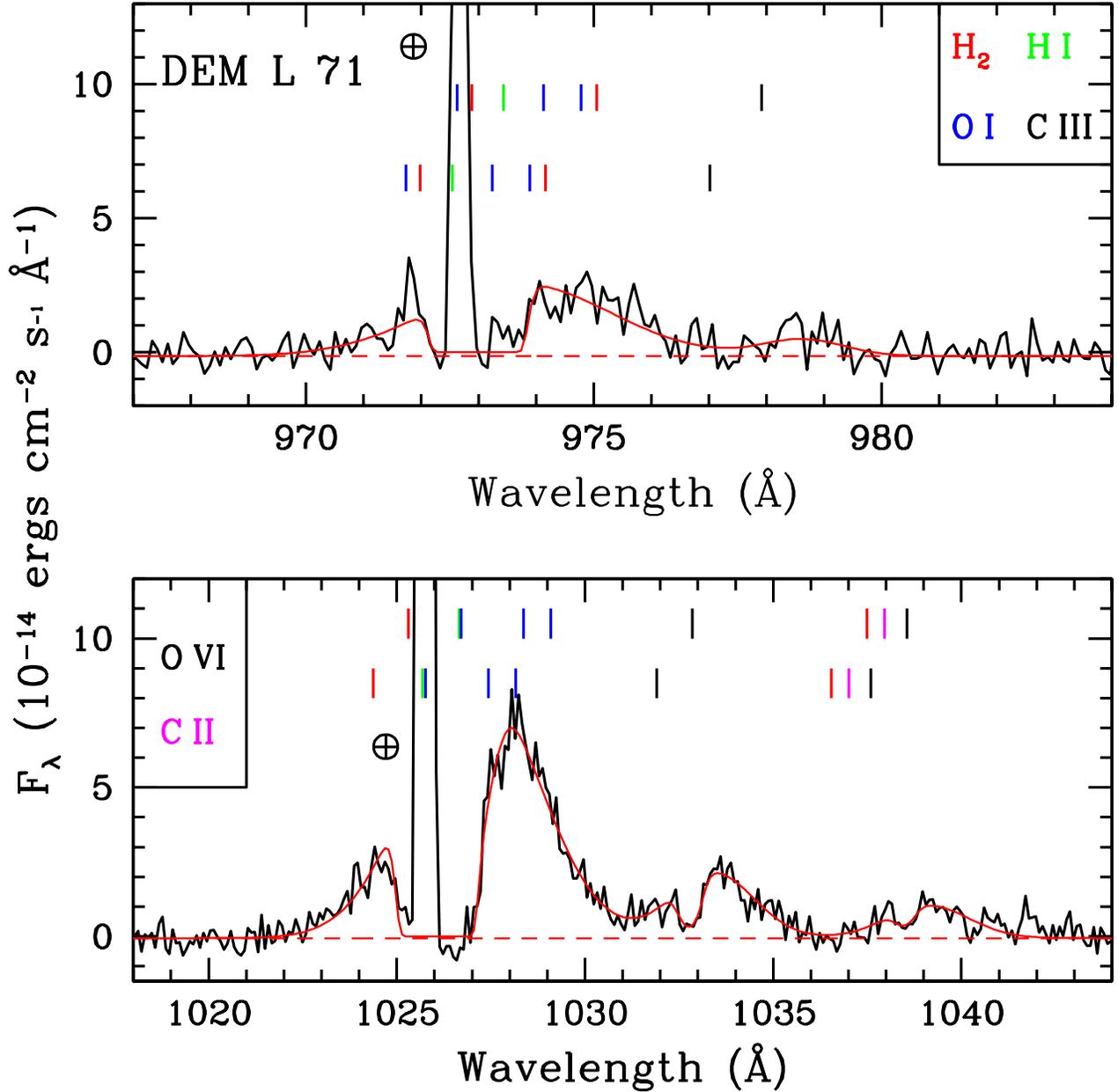}
\caption{Extracted LWRS spectrum of the face-on shock position in \dem71.  The 
best fit line profiles are marked, along with the positions of major interstellar
features from LMC (top tick marks) and Galactic (bottom tick marks) absorption.  The model components
included in the fit are shown in Table~2.   }
\end{figure}

\begin{figure}
\plotone{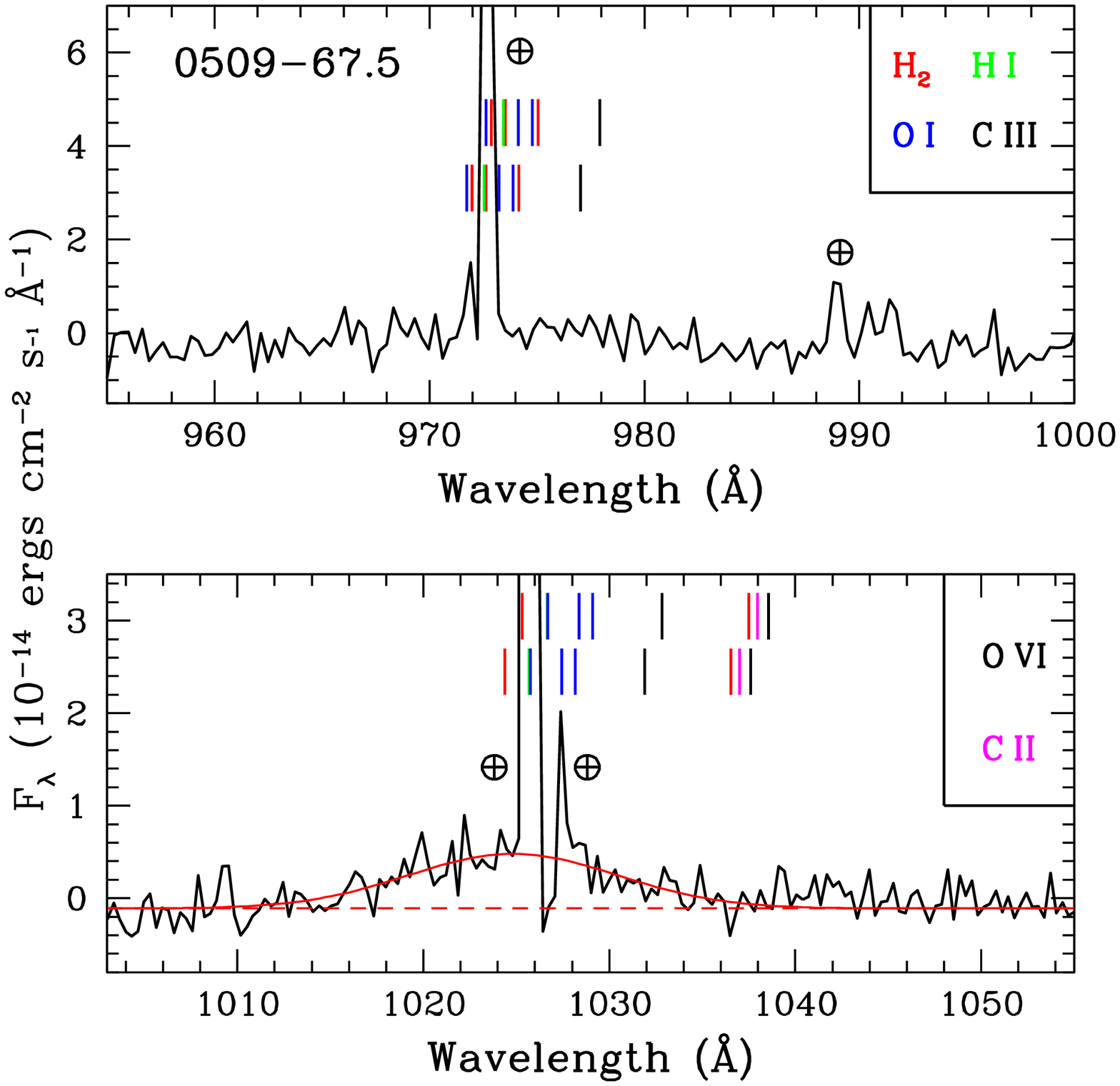}
\caption{Extracted LWRS spectrum of \onine.  The   
best fit line profiles are marked, along with the positions of major interstellar  
features from LMC (top) and Galactic (bottom) absorption.  The ISM absorption
features have been excluded from the fit.  The model components   
included in the fit are shown in Table~2.  }
\end{figure}

\begin{figure}
\plotone{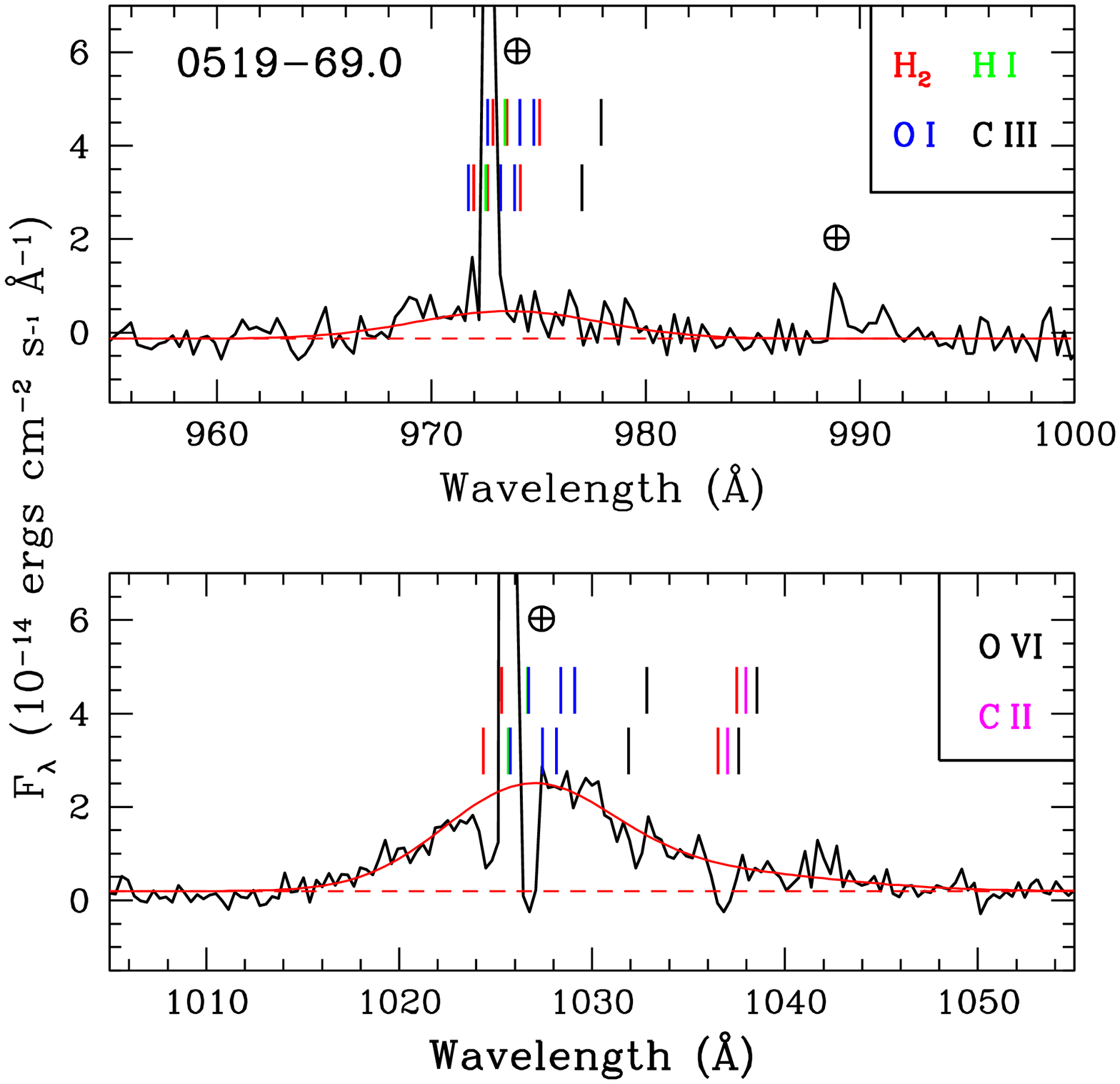}
\caption{Extracted LWRS spectrum of \onenine.  The
best fit line profiles are marked, along with the positions of major interstellar
features from LMC (top) and Galactic (bottom) absorption.  The ISM
absorption features have been excluded from the fit.  The model components   
included in the fit are shown in Table~2.  }
\end{figure}

\begin{figure}
\plotone{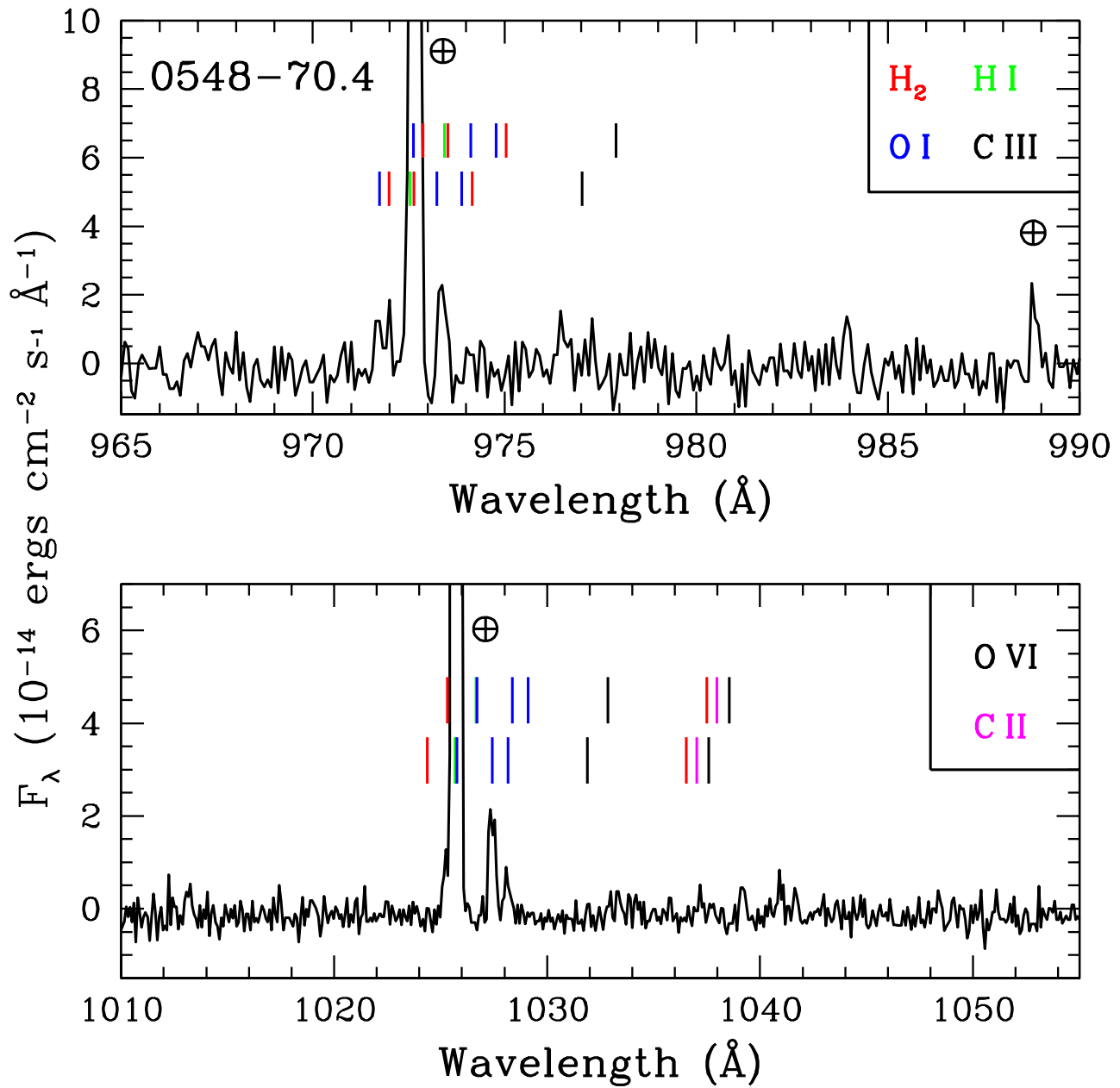}
\caption{Extracted LWRS spectrum of \foureight.  No FUV emission is detected
down to the S/N = 3 sensitivity limit of \fuse\, ($\sim$10$^{-15}$ ergs cm$^{-2}$ s$^{-1}$ \AA$^{-1}$.
at 1032 \AA\, for night only exposures. }
\end{figure}

\begin{figure}
\plotone{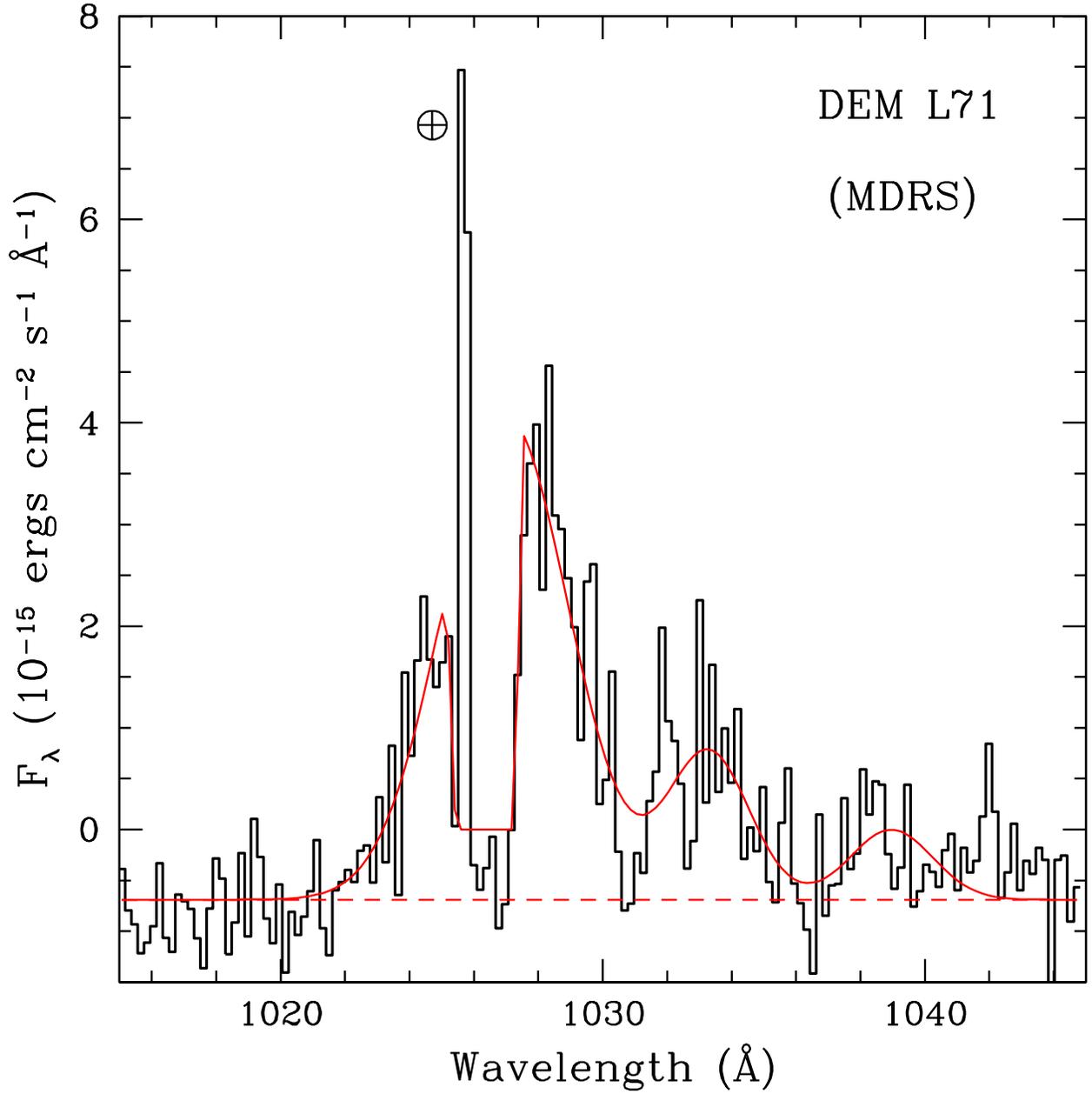}
\caption{The MDRS spectrum of \dem71, acquired from the southern
rim of the SNR.  The fitted emission line components (broad Ly $\beta$, 
O~VI $\lambda\lambda$1032, 1038 and Galactic + LMC H~I absorption are 
marked.  The Ly $\beta$ airglow line (marked by the $\oplus$) is excluded
from the fit.    }
\end{figure}

\begin{figure}
\plotone{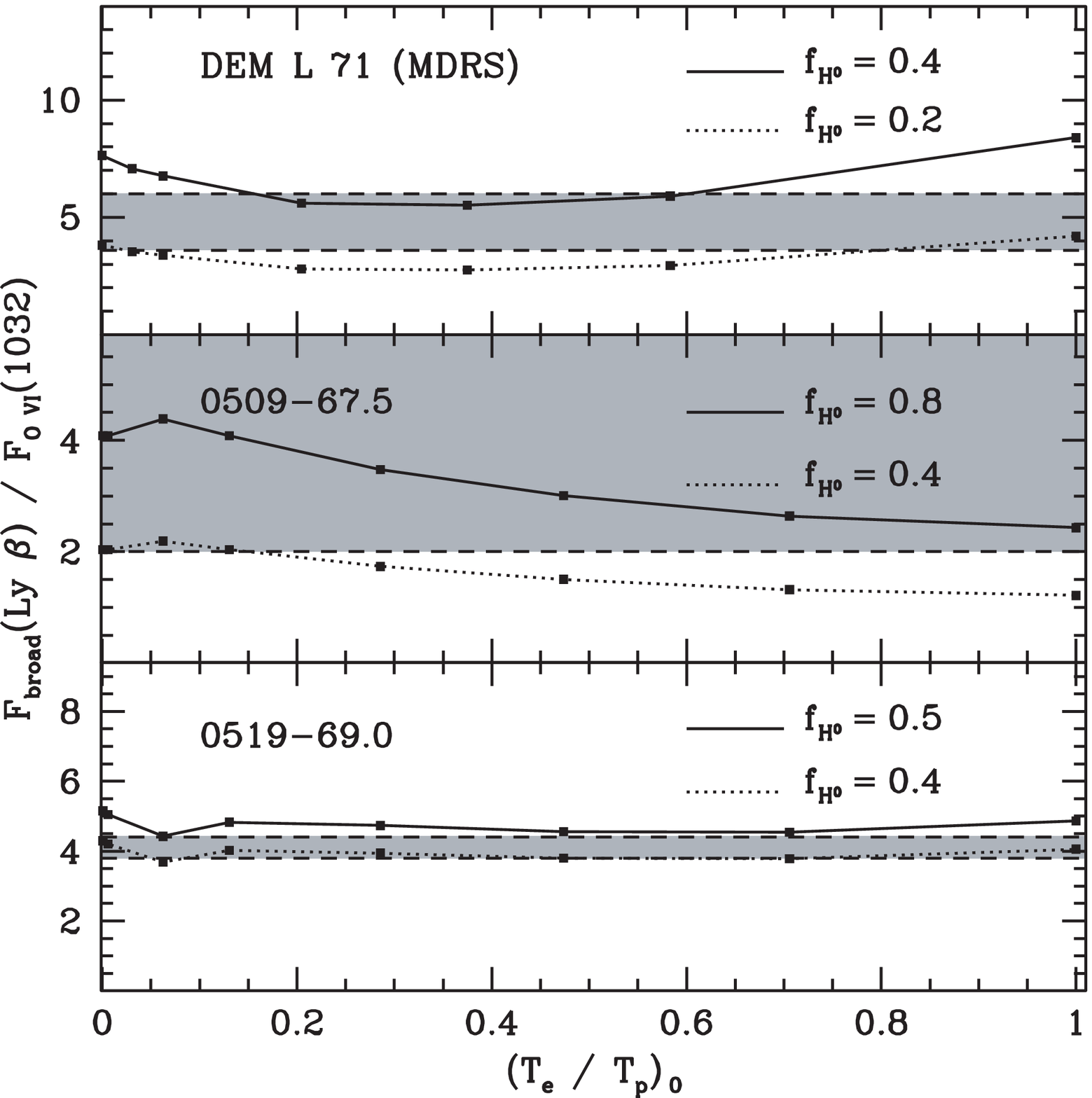}
\caption{The predicted broad \lybovi\, flux ratios, shown as a function of the
assumed electron-proton temperature equilibration for the three FUV spectra
dominated by edge-on blast wave emission.  The measured ranges of the \lybovi\,
ratios are marked by the shaded regions.  The solid and dotted lines show
the predicted flux ratio curves for the maximum and minimum preshock neutral
fractions allowing agreement between the observed and predicted flux ratios.
In the case of \onine\, the maximum preshock neutral fraction (80\%) is 
the largest plausible value given ionization conditions in the warm neutral ISM.
}
\end{figure}

\begin{figure}
\plotone{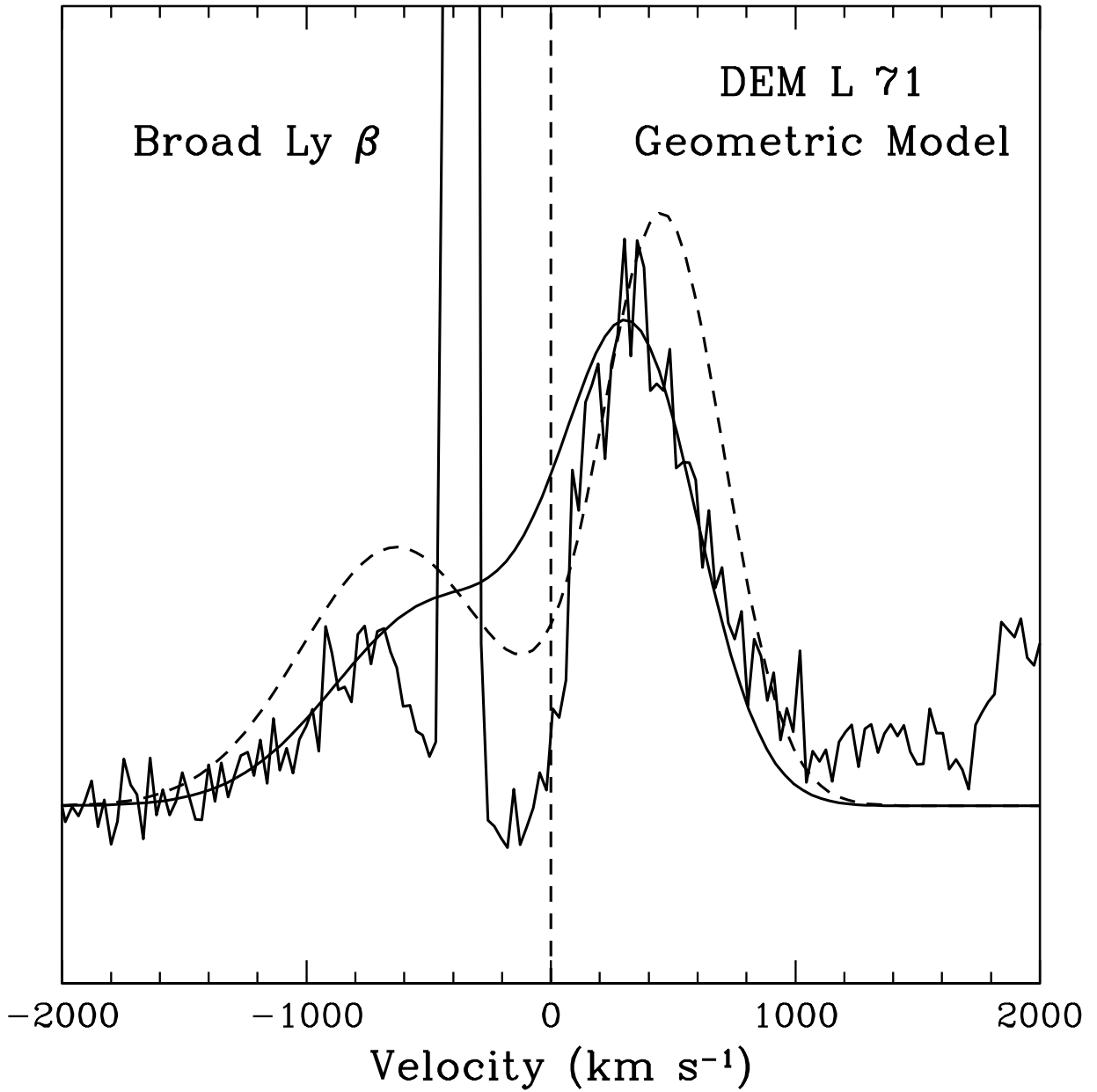}
\caption{The observed and modeled Ly $\beta$ profile from the face-on spectrum
of \dem71\, (Figure~3).  The solid line shows our best approximation to the profile
from our spherical shell models (see text for details).  The dashed line shows our
best approximation when the LWRS aperture (Figure~1) extends only halfway to
the edge of the shell (i.e., no edge-on component is included).  The prominent
absorption feature and Ly $\beta$ airglow line between $-$700 \kms\, and 
+500 \kms\, are excluded from the profile modeling.
 }
\end{figure}

\end{document}